\documentclass[arxiv]{imsart}

\RequirePackage{amsthm,amsmath,amsfonts,amssymb}
\RequirePackage[authoryear]{natbib}%% uncomment this for author-year citations
\RequirePackage[colorlinks,citecolor=blue,urlcolor=blue]{hyperref}
\RequirePackage{graphicx}

\usepackage{multirow}
\usepackage{color}
\usepackage{xcolor}
\usepackage{subcaption}
\usepackage{xr}
\numberwithin{equation}{section} 

\startlocaldefs
\theoremstyle{plain}
\newtheorem{theorem}{Theorem}
\newtheorem{lemma}{Lemma}

\newtheorem{proposition}{Proposition}
\newtheorem{assumption}{Assumption}
\theoremstyle{remark}
\newtheorem*{example}{Example}
\newtheorem*{remark}{Remark}
\def\0{\mathbf{0}}
\def\1{\mathbf{1}}
\def\2{\mathbf{2}}
\def\3{\mathbf{3}}
\def\4{\mathbf{4}}
\def\5{\mathbf{5}}
\def\6{\mathbf{6}}
\def\7{\mathbf{7}}
\def\8{\mathbf{8}}
\def\9{\mathbf{9}}

\def\a{\mathbf{a}}
\def\b{\mathbf{b}}
\def\c{\mathbf{c}}

\def\k{\mathbf{k}}

\def\p{\mathbf{p}}

\def\s{\mathbf{s}}

\def\u{\mathbf{u}}

\def\w{\mathbf{w}}

\def\z{\mathbf{z}}

\def\H{\mathbf{H}}
\def\I{\mathbf{I}}

\def\M{\mathbf{M}}

\def\R{\mathbf{R}}
\def\S{\mathbf{S}}

\def\X{\mathbf{X}}
\def\Y{\mathbf{Y}}

\def\calA{\mathcal{A}}
\def\calB{\mathcal{B}}

\def\calD{\mathcal{D}}

\def\calH{\mathcal{H}}
\def\calI{\mathcal{I}}

\def\calL{\mathcal{L}}

\def\calN{\mathcal{N}}

\def\calP{\mathcal{P}}

\def\calS{\mathcal{S}}
\def\calT{\mathcal{T}}
\def\calU{\mathcal{U}}
\def\calV{\mathcal{V}}
\def\calW{\mathcal{W}}

\def\bbE{\mathbb{E}}

\def\bbI{\mathbb{I}}

\def\bbR{\mathbb{R}}

\def\bfalpha{\boldsymbol \alpha}
\def\bfdelta{\boldsymbol \delta}
\def\bfmu{\boldsymbol \mu}

\def\hd{high dimensional}
\def\tsX{\boldsymbol{\mathsf{X}}}

\def\bfSigma{\boldsymbol \Sigma}

\def\transpose{^{ \mathrm{\scriptscriptstyle T} }}

\def\be{\begin{equation}}
\def\ee{\end{equation}}
\def\bea{\begin{eqnarray}}
\def\eea{\end{eqnarray}}
\def\nn{\nonumber}

\usepackage{color}

\usepackage{ulem}
\let\isout\sout
\renewcommand{\sout}[1]{\ifmmode\text{\isout{\ensuremath{#1}}}\else\isout{#1}\fi}

\newcommand{\add}[1]{#1}
\newcommand{\delete}[1]{}

\newcommand{\mnote}[1]{}
\endlocaldefs

\begin{document}

\begin{frontmatter}
\title{Localization Estimator for High Dimensional Tensor Covariance Matrices}
%\title{A sample article title with some additional note\thanksref{t1}}
\runtitle{localization covariance estimator}
%\thankstext{T1}{A sample additional note to the title.}

\begin{aug}
%%%%%%%%%%%%%%%%%%%%%%%%%%%%%%%%%%%%%%%%%%%%%%%
%% Only one address is permitted per author. %%
%% Only division, organization and e-mail is %%
%% included in the address.                  %%
%% Additional information can be included in %%
\author[A]{\fnms{Hao-Xuan}~\snm{Sun}\ead[label=e1]{hxsun@hit.edu.cn}\orcid{0000-0001-5062-3968}},
\author[B]{\fnms{Song Xi}~\snm{Chen}\ead[label=e2]{songxichen@pku.edu.cn}\orcid{0000-0002-2338-0873}}
\and
\author[C]{\fnms{Yumou}~\snm{Qiu}\ead[label=e3]{qiuyumou@math.pku.edu.cn}\orcid{0000-0003-4846-1263}}
%%%%%%%%%%%%%%%%%%%%%%%%%%%%%%%%%%%%%%%%%%%%%%
%% Addresses                                %%
%%%%%%%%%%%%%%%%%%%%%%%%%%%%%%%%%%%%%%%%%%%%%%
\address[A]{School of Mathematics,
Harbin Institute of Technology\printead[presep={,\ }]{e1}}

\address[B]{Department of Statistics and Data Science,
Tsinghua University\printead[presep={,\ }]{e2}}

\address[C]{School of Mathematical Sciences and Center for Statistical Science,
Peking University\printead[presep={,\ }]{e3}}
\end{aug}

\begin{abstract}
   This paper considers covariance matrix estimation of tensor data under high dimensionality. A multi-bandable covariance class is established to accommodate the need for complex covariance structures of multi-layer lattices and general covariance decay patterns. We propose a high dimensional covariance localization estimator for tensor data, which regulates the sample covariance matrix through a localization function. The statistical properties of the proposed estimator are studied by deriving the minimax rates of convergence under the spectral and the Frobenius norms. Numerical experiments and real data analysis on ocean eddy data are carried out to illustrate the utility of the proposed method in practice.
\end{abstract}

\begin{keyword}[class=MSC]
\kwd[Primary ]{62H12}
\kwd[; secondary ]{62C20}
\end{keyword}

\begin{keyword}
\kwd{Covariance matrix Estimation}
\kwd{High dimensionality}
\kwd{Localization}
\kwd{Minimax rate} 
\kwd{Tensor Data}
\end{keyword}

\end{frontmatter}
%%%%%%%%%%%%%%%%%%%%%%%%%%%%%%%%%%%%%%%%%%%%%%
%% Please use \tableofcontents for articles %%
%% with 50 pages and more                   %%
%%%%%%%%%%%%%%%%%%%%%%%%%%%%%%%%%%%%%%%%%%%%%%
%\tableofcontents

\section{Introduction}

Estimation of covariance matrices is a basic task in statistics as the covariance matrices and their estimates play a key role in many multivariate statistical procedures. For the fixed dimensional case, the estimation quality of the sample covariance matrix can be assured by the conventional multivariate analysis.  However, for \hd\ problems, the consistency of the sample covariance matrix is no longer guaranteed \citep{Bai1993, Bai1998, Johnstone2001}. The last two decays have seen consistent \hd\ covariance estimators being proposed, which include the banding and thresholding estimators proposed by \cite{Bickel2008} and \cite{Bickel2008b}, the tapering estimator proposed by \cite{Cai2010}, the adaptive thresholding method in \cite{Cai2011}, the block thresholding method in \cite{Cai2012}, and the separately banding and tapering estimators in \cite{Zhang2023}.

The \hd\ banding and tapering estimators are for covariances with the so-called bandable covariance structure, given in the pioneering work \cite{Bickel2008}, which very much reflected a univariate \hd\ problem where the components of the data vector follow a natural ordering with respect to their dependence (covariance). One such example is in modeling a climate variable on a fixed latitude over a longitude range as is the case for Lorenz models, where the high dimensionality is created by finer resolution observations of the climate variable \citep{Sun2024}; another is for genetic observations collected on a chromosome. The separately bandable covariance class adopted in \cite{Zhang2023} is a bivariate extension of the univariate \hd\ system, which has two directions (for instance latitude and longitude) that govern the covariance ordering of the random components of the \hd\ observations. Despite their success in univariate and bivariate settings, the aforementioned \hd\ covariance estimators encounter limitations for tensor data, in which case the structural assumptions of the bandable or separably bandable covariances may be overly restrictive for accurately capturing complex dependence patterns embedded in tensor data driven by multi-sources.

Tensor data, represented as multi-layer arrays, can exhibit complex dependencies arising from their underlying multi-source drivers, where heterogeneity may exist across different dimensions of the tensor. In many scientific disciplines, \hd\ tensor data are increasingly collected from studies when the target observations are generated from multiple sources. An example is in oceanic studies on certain variables, say sea temperature or salinity, where tensor data with respect to latitude, longitude and height and depth, are collected, and the high dimensionality is created via high spatial resolution of observations due to an ever increasing measuring capability.

This study aims at developing a \hd\ covariance matrix estimation method for tensor data. Without loss of generality, we assume that the tensor data are embedded over a multi-layer lattice. We propose a multi-bandable covariance class to accommodate the complex covariance structures of multi-layer tensor data, upon which consistent estimation of the covariance matrix is possible for tensor data. The proposed covariance class is formally defined via a covariance decay function that governs the covariance structure relative to the underlying lattice positional relationship. Such construction can permit general covariance decay patterns, including the polynomial and exponential decays as functions of the inter-grid point distances and the heterogeneous setting where the covariances possess different decay rates in different layers. Therefore, the proposed covariance class extends the existing bandable \cite{Bickel2008} and separably bandable \cite{Zhang2023} covariance classes to a comprehensive framework that incorporates flexible covariance behaviors, while offering richer  bandable-in-bandable and multi-bandable covariance structures that may arise in the tensor settings.

We propose a \hd\ covariance localization estimator for tensor data that subscribes to the multi-bandable covariance structure. The proposed method implements regularization on the sample covariance estimation through a localization function with scaling parameters, which can permit non-linearly decayed weight functions and heterogeneous scaling parameters to better suit the complex covariance structures in the multi-bandable covariance class. Theoretical analyses establish the consistency of the proposed estimator under both the spectral and the Frobenius norms, with the minimax optimal convergence rates being achieved with the optimal choice of the scaling parameters. These results significantly generalize the existing results in \hd\ settings in \cite{Bickel2008} and \cite{Cai2010} largely for basically univariate but high dimensional tensor data. Our analyses also extend the results of \cite{Zhang2023} for bivariate tensor data to tensor data with the number of layers larger than two and to allow non-separable covariance structure. Numerical experiments and a case study on an ocean eddy tensor data confirmed the performance of the proposed method.

The rest of the paper is organized as follows. We first present the \hd\ covariance estimation problem setting for tensor data through an ocean eddy dataset in Section \ref{sec:setting}. A multi-bandable covariance class is proposed in Section \ref{sec:multi-bandable-class}, followed by a \hd\ covariance localization estimator in Section \ref{sec:local-est}. The consistency and the minimax optimal convergence rate of the proposed estimation are established in Section \ref{sec:main-result} for both the spectral and the Frobenius norms. Sections \ref{sec:simulation} and \ref{sec:case} report the simulation studies and real data analysis, respectively. The conclusion is finally provided in Section \ref{sec:conclusion}. The technical proofs and additional theoretical and numerical results are presented in supplementary material (SM).

\section{Preliminaries}
\label{sec:setting}

Throughout this paper, we use italic letters $a, b$ for scalars and bold Roman letters $\a, \b$ for vectors and matrices. We use $\0_p$ and $\1_p$ to denote the $p$-dimensional vectors with all elements being $0$ or $1$, respectively. We write $a\asymp b$ if there are positive constants $C_1$ and $C_2$ such that $C_1\le a/b \le C_2$. For a vector $\a$, we use $\|\a\|$ to denote its Euclidean norm. For a matrix $\M$, we use $\|\M\|$ and $\|\M\|_F$ to denote the spectral and the Frobenius norm, respectively.

As we aim at developing consistent covariance estimators for tensor data, the following notations are introduced for the multi-bandable covariance class. Specifically, the tensor data in this work are assumed to be organized over a multi-layer lattice. Let $\tsX:=\{X(\s)\}_{\s\in\calS_d(\p)}$ be tensor data sampled from a $d$-order lattice
\begin{equation*}
    \calS_d(\p)=\{1,2,\dots,p_1\}\times\{1,2,\dots,p_2\}\times\dots\times\{1,2,\dots,p_d\},
\end{equation*}
where $\p=(p_1,\dots,p_d)\transpose$ are the dimensions of the tensor elements in $d$ directions. We consider in this work a regime where $d$ is fixed but $p_\ell\to\infty$ for all $\ell=1,\dots,d$. Denote by $\{\s_1,\s_2,\dots,\s_p\}$ an arbitrary arrangement of the total $p=\prod_{\ell=1}^dp_\ell$ entries in $\calS_d(\p)$, where $\s_i=(s_{i1},\dots,s_{id})\transpose$ denotes the coordinate of the $i$th entry. Then, a vectorization of a tensor data $\tsX$ according to the arrangement $\{\s_1,\s_2,\dots,\s_{p}\}$ is $\X=(X(\s_1),\dots,X(\s_p))\transpose$. The lattice $\calS_d(\p)$ is introduced to represent a general class of data that can be mapped into some ordered tensor elements. For example, $d=1$ represents an univariate \hd\ problem which is the situation largely studied in \cite{Bickel2008} and $d=2$ corresponds to matrix data \cite{Zhang2023}. While $d>1$ generalizes to tensors of arbitrary order, which are also common cases in many scientific research, for instance, the oceanic data which will be presented shortly corresponds to $d=3$. 

Suppose $\tsX_1,\tsX_2,\dots,\tsX_n$ are independent and identically distributed copies of a random tensor $\tsX$, with their vectorizations being $p$-dimensional random vectors $\X_1,\X_2,\dots,\X_n$ with mean vector $\bfmu$ and covariance matrix $\bfSigma=[\sigma_{ij}]_{p\times p}$. Specifically, $\X_i=(X_i(\s_1), \allowbreak X_i(\s_2),\dots,X_i(\s_{p}))\transpose$ denotes the $i$th element of the underlying random tensor on the entire lattice $\calS_d(\p)$. We are interested in the estimation of the covariance matrix $\bfSigma$.

A natural estimator of the covariance matrix $\bfSigma$ is the sample covariance matrix
\begin{eqnarray}
    \label{eq:sample-covariance}
    \S_{n} = \frac{1}{n-1}\sum_{m=1}^n\big(\X_m-\bar{\X}\big)\big(\X_m-\bar{\X}\big)\transpose=[\hat{\sigma}_{ij}]_{p\times p},
\end{eqnarray}
where $\bar{\X}=n^{-1}\sum_{m=1}^n\X_m$. For the ``large $p$ small $n$" situations, the sample covariance matrix is no longer consistent with $\bfSigma$ \citep{Muirhead1987, Bai1993, Bai1998} and the solution is to take into account the underlying covariance structure.

Previous studies constructed several consistent covariance matrix estimators under certain covariance classes to cope with the high dimensionality. 
\cite{Bickel2008} considered a banding estimator
\begin{equation}
    \label{eq:banding-estimator}
    \calB_k(\S_n)=[\hat{\sigma}_{ij}\bbI\{|i-j|\le k\}]
\end{equation}
at a banding width $k$ for a bandable covariance class
\begin{eqnarray}\label{eq:bandable-class-BL}
    \calU_1(\alpha,\epsilon,C) = \Big\{\bfSigma=[\sigma_{ij}]_{p\times p}: &\text{(i)}& \max_j\sum_{|i-j|>k}|\sigma_{ij}|\le Ck^{-\alpha} \hbox{ for all } k>0, \nn \\
    &\text{(ii)}& 0\le\lambda_{\min}(\bfSigma)\le \lambda_{\max}(\bfSigma)\le \epsilon^{-1}\Big\},
\end{eqnarray}
where $\alpha$, $C$ and $\epsilon$ are some positive constants and \add{$\lambda_{\min}(\M)$ and} $\lambda_{\max}(\M)$ denote the \add{minimum and} maximum eigenvalues of a matrix $\M$\add{, respectively}. \cite{Bickel2008} also imposes a minimum eigenvalue condition on the bandable covariance class for estimating the precision matrix. However, this condition is not necessary for the covariance estimation. They established that by choosing $k\asymp(n^{-1}\log p)^{-1/(2\alpha+2)}$ the estimation error of the banding estimator satisfies 
\begin{equation}
    \label{eq:convergence-rate-BL}
    \|\calB_k(\S_n)-\bfSigma\|^2=O_p\{(n^{-1}\log p)^{2\alpha/(2\alpha+2)}\}.
\end{equation}
\cite{Cai2010} introduced a tapering estimator
\begin{eqnarray}
    \label{eq:linear-taper-estimator}
    \calT_{k}(\S_n)=\big[\hat{\sigma}_{ij}\varphi(|i-j|;k/2,k)\big]_{p\times p}
\end{eqnarray}
where 
\begin{eqnarray}\label{eq:linear-taper}
   \varphi(z;k_a,k_b) = \bbI\big\{z\le k_a\big\}+\frac{k_b-z}{k_b-k_a}\bbI\big\{k_a<z\le k_b\big\}
\end{eqnarray} 
is a tapering function. The estimator was designed for both $\calU_1(\alpha,\epsilon,C)$ and the following covariance class similar to $\calU_1(\alpha,\epsilon,C)$ 
\begin{eqnarray}\label{eq:bandable-class-CZZ-2}
    \calU_2(\alpha,\epsilon,C) = \Big\{\bfSigma=[\sigma_{ij}]_{p\times p}: &\text{(i)}& |\sigma_{ij}|\le C|i-j|^{-\alpha-1} \hbox{ for all } i\neq j,\\
    &\text{(ii)}& 0\le\lambda_{\min}(\bfSigma)\le\lambda_{\max}(\bfSigma)\le \epsilon^{-1}\Big\}, \nonumber
\end{eqnarray}
which is more restrictive on the off-diagonal elements. It is shown that the tapering estimator can achieve the following minimax optimal rates of convergence
\begin{equation}
    \label{eq:convergence-rate-CZZ}
    \inf_{\hat\bfSigma}\sup_{\calU_1(\alpha,\epsilon,C)}\bbE\|\hat\bfSigma-\bfSigma\|^2\asymp\min\Big\{n^{-\frac{2\alpha}{2\alpha+1}}+\frac{\log p}{n},\frac{p}{n}\Big\}
\end{equation}
with $k\asymp \min\{n^{1/(2\alpha+1)},p\}$ and
\begin{equation*}
    \inf_{\hat\bfSigma}\sup_{\calU_2(\alpha,\epsilon,C)}p^{-1}\bbE\|\hat\bfSigma-\bfSigma\|_F^2\asymp\min\Big\{n^{-\frac{2\alpha+1}{2\alpha+2}},\frac{p}{n}\Big\}
\end{equation*}
with $k\asymp \min\{n^{1/(2\alpha+2)},p\}$.

From another point of view, both bandable classes $\calU_1(\alpha,\epsilon,C)$ and $\calU_2(\alpha,\epsilon,C)$ regulate the magnitude of the covariance elements in terms of $|i-j|$,  which were largely designed for univariate tensor data with $d=1$.

\cite{Zhang2023} extended the bandable covariance classes $\calU_1(\alpha,\epsilon,C)$ and  $\calU_2(\alpha,\epsilon,C)$ to matrix data ($d=2$) by assuming separability in the covariance between the two coordinate directions. Specifically, denote by $\otimes$ the Kronecker product and $\text{vec}(\cdot)$ the vectorization operator that sequentially stacks the columns of a matrix into a vector. Supposed that the matrix data $\X\in\bbR^{p_1\times p_2}$ follows a matrix normal distribution such that $\text{vec}(\X)\sim\calN(\text{vec}(\M),\bfSigma_2\otimes\bfSigma_1)$ for a mean matrix $\M\in\bbR^{p_1\times p_2}$ and the covariance matrices $\bfSigma$ being within a separably bandable covariance class 
\begin{eqnarray}
    \label{eq:bandable-class-ZSK}
    \calU_{3,q}(\alpha_1,\alpha_2,\epsilon,C)=\big\{\bfSigma=\bfSigma_2\otimes \bfSigma_1: && \bfSigma_1\in\calU_q(\alpha_1,\epsilon,C), \bfSigma_2\in\calU_q(\alpha_2,\epsilon,C) \\
    && \hbox{ and } \min\{\lambda_{\min}(\bfSigma_1),\lambda_{\min}(\bfSigma_2)\}> \epsilon^{-1}\big\}
\end{eqnarray}
for either $q=1$ and $2$. \delete{, where $\lambda_{\min}(\M)$ represents the minimum eigenvalue of a matrix $\M$.} They then proposed a separably banding or tapering estimator $\hat\bfSigma_2(k_2) \otimes \hat\bfSigma_1(k_1)$ with  
\begin{eqnarray}\label{eq:ZSK-approximation}
    \big(\hat\bfSigma_1(k_1), \hat\bfSigma_2(k_2)\big)=\underset{\bfSigma_1,\bfSigma_2}{\arg\min}\|\tilde{\bfSigma}(k_1,k_2)-\bfSigma_2\otimes \bfSigma_1\|_F^2,
\end{eqnarray}
where $k_1$ and $k_2$ are banding width parameters corresponding to the row and column of the matrix data, $\tilde{\bfSigma}(k_1,k_2)$ is a doubly banding or tapering estimator that regularizes the sample covariance as $\S_n\circ \{\calB_{k_2}(\1_{p_2}\1_{p_2}\transpose)\otimes \calB_{k_1}(\1_{p_1}\1_{p_1}\transpose)\}$ or $\S_n\circ \{\calT_{k_2}(\1_{p_2}\1_{p_2}\transpose)\otimes \calT_{k_1}(\1_{p_1}\1_{p_1}\transpose)\}$, respectively, with $\circ$ being the elementwise product. For these estimators, they derived the upper bound on the standardized error of estimation  
\begin{equation*}
    \label{eq:upper-ZSK}
    \bbE\Big(\frac{\|\hat\bfSigma_2(k_2)\otimes\hat\bfSigma_1(k_1)-\bfSigma\|_F^2}{p_1p_2}\Big)=O_p\Big\{\min\Big(\frac{k_1k_2}{n},\frac{p_1k_1^2}{p_2n^2}+\frac{p_2k_2^2}{p_1n^2}\Big)+k_1^{-\tilde\alpha_1}+k_2^{-\tilde\alpha_2}\Big\},
\end{equation*}
for $k_1$ and $k_2$ satisfying $k_1<p_1$, $k_2<p_2$ and $p_1k_1+p_2k_2> Cn$ for a positive constant $C$ where $\tilde\alpha_\ell=2\alpha_\ell$ for $\bfSigma\in\calU_{3,1}(\alpha_1,\alpha_2,\epsilon,C)$ and $\tilde\alpha_\ell=2\alpha_\ell+1$ for $\bfSigma\in\calU_{3,2}(\alpha_1,\alpha_2,\epsilon,C)$. \cite{Zhang2023} also provided the minimax lower bound for the estimation error under the Frobenius norm. 

The aforementioned covariance classes can be too crude to reflect the underlying covariance structure encountered for general tensor data with $d \ge 3$. For data from a multi-order lattice with $d\ge 2$, the measurement $|i-j|$ in the bandable covariance classes $\calU_{1}(\alpha_,\epsilon,C)$ and $\calU_{2}(\alpha_,\epsilon,C)$ may be overly simple, as the detailed covariance information especially, the intricate cross-dimensional covariances, can be richer than what the covariance class $\calU_1(\alpha,\epsilon,C)$ or $\calU_2(\alpha,\epsilon,C)$ can offer. Although the separably bandable covariance class $\calU_{3,q}(\alpha_1,\alpha_2,\epsilon,C)$ in \eqref{eq:bandable-class-ZSK} has gone beyond $\calU_{1}(\alpha_,\epsilon,C)$ and $\calU_{2}(\alpha_,\epsilon,C)$ to adapt to the matrix data, the separability assumption plays a key role and can be restrictive for situations where the separable covariance may not be valid. For example, \cite{Daley2001} considered the nonseparable error correlation in an atmospheric variational data assimilation system, while \cite{Hakim2005} revealed the nonseparability nature between the horizontal and the vertical structure of forecast and analysis error {covariance matrix} for mid-latitude atmospheric data assimilation over the western north Pacific. Specific nonseparable spatial-temporal covariance structures can also be found in \cite{Cressie1999} and \cite{Guttorp2013}. In the meanwhile, a more challenging case is when the data are sampled from a multi-layer lattice with $d\ge 3$, which is commonly encountered in geophysical research.

\begin{figure}[ht!]
    \centering
    \begin{subfigure}[b]{\textwidth}
        \caption{Sea level anomaly (left) and salinity field (right) of an eddy}
        \includegraphics[width=\textwidth]{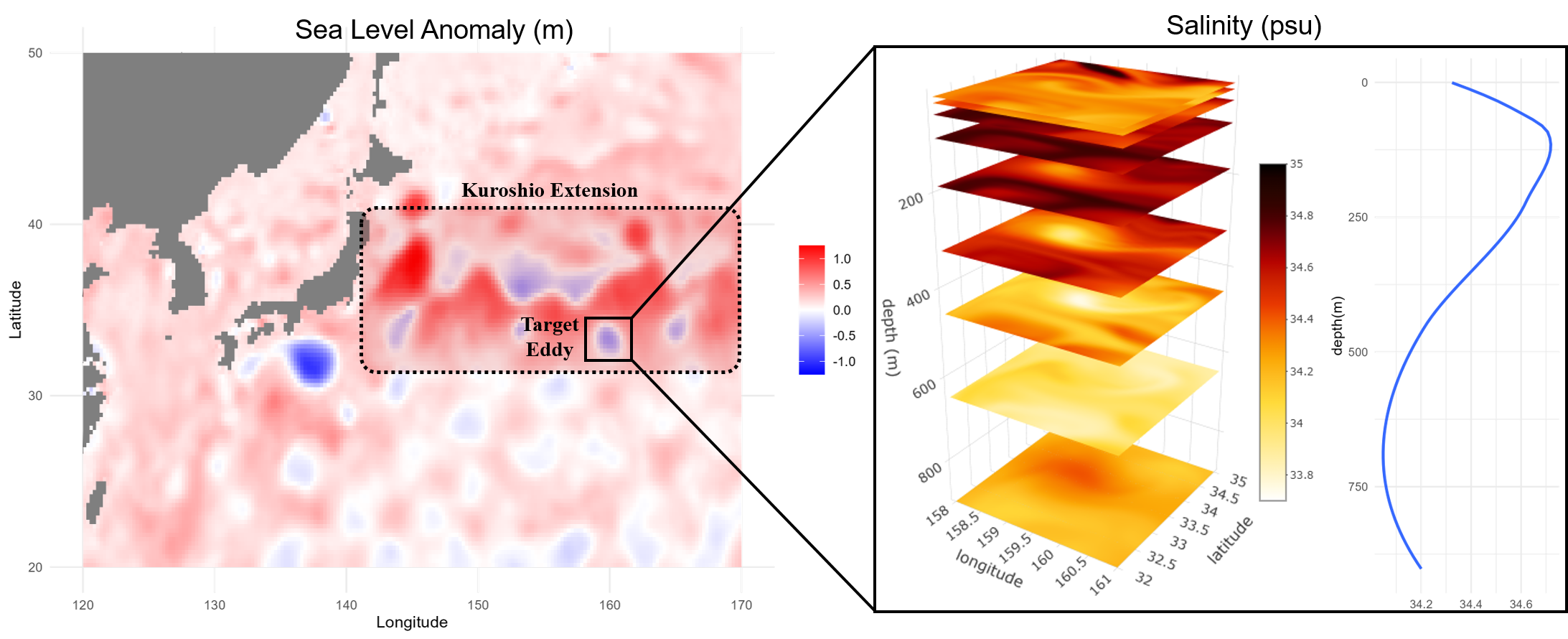}
        \label{fig:Eddy-1}
    \end{subfigure}
    \begin{subfigure}[b]{\textwidth}
        \caption{Correlation matrix and insert of daily salinity changes}
        \includegraphics[width=\textwidth]{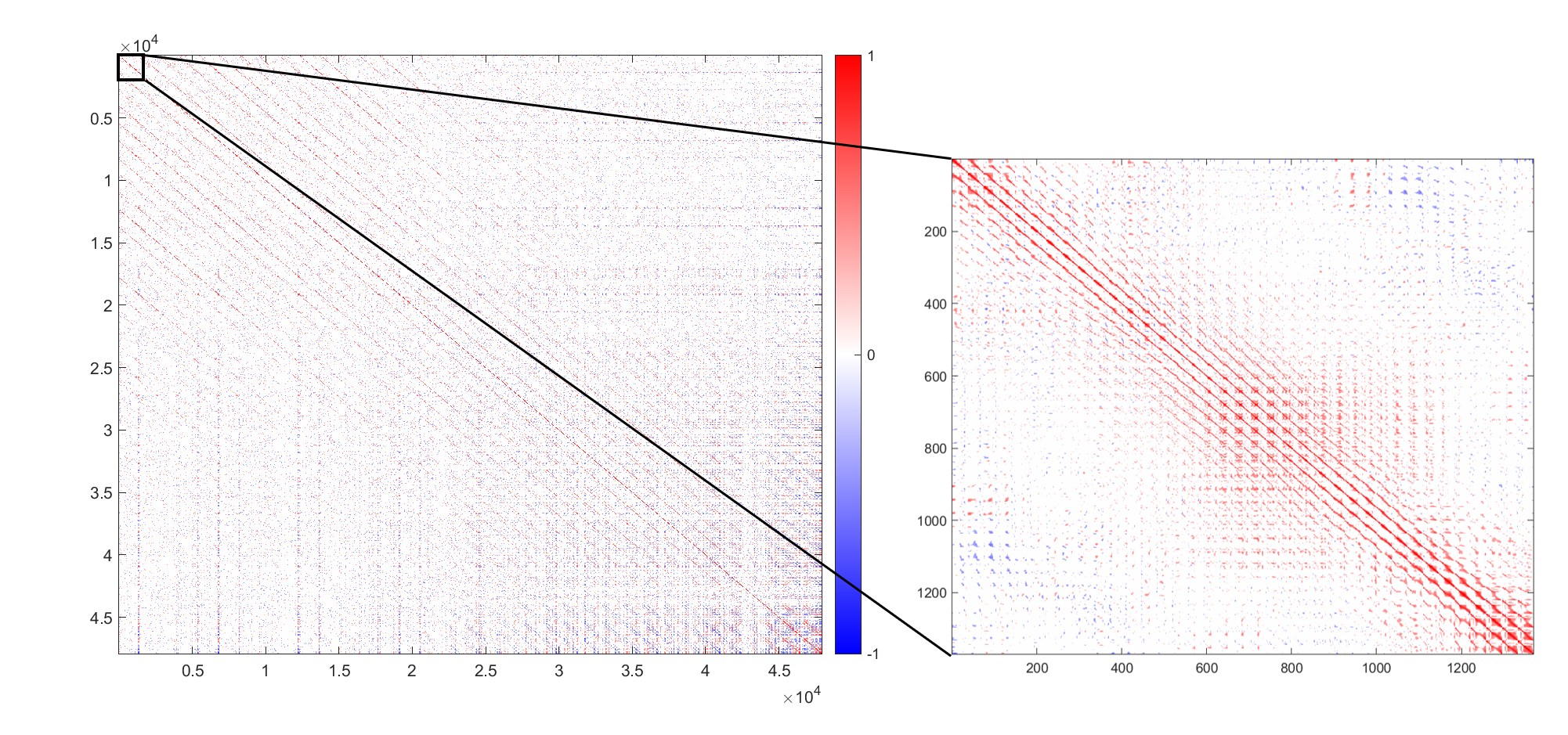}
        \label{fig:Eddy-2}
    \end{subfigure}
    \caption{(a) Sea level anomaly on September 1st, 2024 that displays an ocean eddy at 32$^\circ$N-35$^\circ$N and 158$^\circ$E-161$^\circ$E. %within Kuroshio Extension
    (left) and the trivariate salinity field in the practical salinity unit (psu) in $10$ out of $35$ total layers and the average salinity with respect to the depth (right); (b) the sample correlation matrix of daily salinity changes from July 20th to September 12th 2024 at $47915$ grids with the $1/12^\circ\times 1/12^\circ$ spatial resolution and $35$ layers in depth (left), and at the $1369$ grids of the first layer at $0.5\mathrm{m}$ depth (right).}
    \label{fig:Eddy}
\end{figure}

We present a \hd\ ocean eddy dataset which constitute a tensor data with $d=3$, which has motivated our study. The data were the average reanalysis salinity field associated with an ocean eddy in the western Pacific (32$^\circ$N-35$^\circ$N and 158$^\circ$E-161$^\circ$E) from July 20th to September 12th, 2024 with spatial resolution of $1/12^\circ$. We re-centered the eddy with respect to its center each day to make it within a $3^\circ\times 3^\circ$ region and $1\mathrm{km}$ in depth. The eddy center was calculated as the location with the minimum sea level anomaly, which is available each day via satellite observations. Each day was treated as a replication, which led to $54$ observations of the daily salinity changes over $p=47915$ grids, which were evenly distributed in $37\times 37$ longitude-latitude grids, coupled with $35$ vertical layers that were gradually thinning out from $0.5\mathrm{m}$ resolution to $1000\mathrm{m}$.

The trivariate tensor grids were vectorized in the order of the longitude, the latitude and the depth. To be specific, the depth from $0.5\mathrm{m}$ to $1000\mathrm{m}$ changed the slowest from the $1$st to the $47915$th grids, the latitudes were then arranged {from low latitude to high latitude} for each depth while the longitude changes the fastest {from west to east} for each latitude and each depth. Figure \ref{fig:Eddy-2} displays the sample correlation matrix of daily salinity changes in the target area, with an insert corresponding to the correlation in the sea surface.

Although such a correlation matrix displayed a superficial resemblance to the overall bandable structure defined in \cite{Bickel2008}, a closer examination in Section \ref{sec:main-result} reveals its deviation from the bandable class \eqref{eq:bandable-class-BL}. Indeed, it possessed a finer \textit{bandable-in-bandable} form, that offers a richer covariance structure than that offered by the existing bandable class. Hence, the bandable covariance class $\calU_1(\alpha,\epsilon,C)$ in \cite{Bickel2008} would not be detailed enough, as it is much based on the univariate \hd\ problems, for instance, meteorological observations at a fixed latitude or longitude, or genetic observations over a chromosome. Furthermore, the trivariate ocean tensor data may not be adequately captured by the separably bandable covariance class in the spirit of  $\calU_{3,q}(\alpha_1,\alpha_2,\epsilon,C)$ for matrix data. These motivate us to consider the procedure and property of the covariance matrix estimation of the general cases of the $d$-order lattice $\calS_d(\p)$ to accommodate the more complex covariance structure encountered in various statistical applications. For this purpose, a universal class of the covariance matrix and the corresponding estimators shall be developed, which is the focus of the next section. 

\section{Methology}
\label{sec:method}
We have mentioned above that the existing bandable covariance class lacks details to model rich covariance matrices in tensor data. The ocean eddy example in Figure \ref{fig:Eddy} displays a bandable-in-bandble structure, that is, the bandable structures within the banded areas on either side of the main diagonal. The key idea of capturing such characters is to properly use the location information of each grid. For this purpose, we first introduce a multi-bandable covariance class that is suitable for tensor data. A localization estimator is then proposed to provide the covariance matrix estimation under high dimensionality.

\subsection{Multi-bandable Covariance Class}
\label{sec:multi-bandable-class}
We start by introducing some notations to describe the relative position between any two grids in a $d$-order lattice $\calS_d(\p)$, which is aimed to extend the exclusion $|i-j|>k$ used in the bandable covariance class $\calU_1(\alpha,\epsilon,C)$. For the tensor data $\tsX$ sampled from $\calS_d(\p)$ with their vectorizations being $\X=(X(\s_1),X(\s_2),\dots,X(\s_p))\transpose$, we define an absolute difference between two coordinates $\s_i$ and $\s_j$ as 
\begin{eqnarray}\label{eq:absolute-diff}
    \bfdelta_{ij}:=(\delta_{ij1},\delta_{ij2},\dots,\delta_{ijd})\transpose := (|s_{i1}-s_{j1}|,|s_{i2}-s_{j2}|,\dots,|s_{id}-s_{jd}|)\transpose.
\end{eqnarray}
Such a definition allows for different scales and dimensions $p_\ell$ in different coordinate directions of $\calS_d(\p)$. Specifically, the distance of the two coordinates $\s_i$ and $\s_j$ can be represented by some function of $\bfdelta_{ij}$, for example, by $\|\bfdelta_{ij}\|$ for the $L_2$ distance. To impose restriction on the covariance between two grids that are separated apart by some $d$-dimensional vector $\k=(k_1,k_2,\dots,k_d)$ in the set $\calS_d(\p)$, we define $\calH_d(\k)$ as a $d$-dimensional hyper-rectangle region with the side lengths being $\k$, that is
\begin{eqnarray}
    \label{eq:hyper-rectangle}
    \calH_d(\k)=\{\k'=(k_1',k_2',\dots,k_d'):0\le k_\ell'\le k_\ell \hbox{ for } \ell=1,\dots,d\}, 
\end{eqnarray} 
which will be called the preserved $k$-zone. From another point of view, the collection $\calH_d(\k)$ comprises all the absolute coordinate differences that the $\ell$th direction difference between two grids $\s_i$ and $\s_j$ satisfies $\delta_{ij\ell}\le k_\ell$ for all $\ell=1,\dots,d$. This construction generalizes the univariate $|i-j|>k$ used in $\calU_1(\alpha,\epsilon,C)$ in the multi-order lattices, to $\bfdelta_{ij}\notin\calH_d(\k)$ indicating separation of at least $k_\ell$ in the $\ell$th direction between any $\s_i$ and $\s_j$.

\begin{figure}[ht!]
    \centering
    \includegraphics[width = \textwidth]{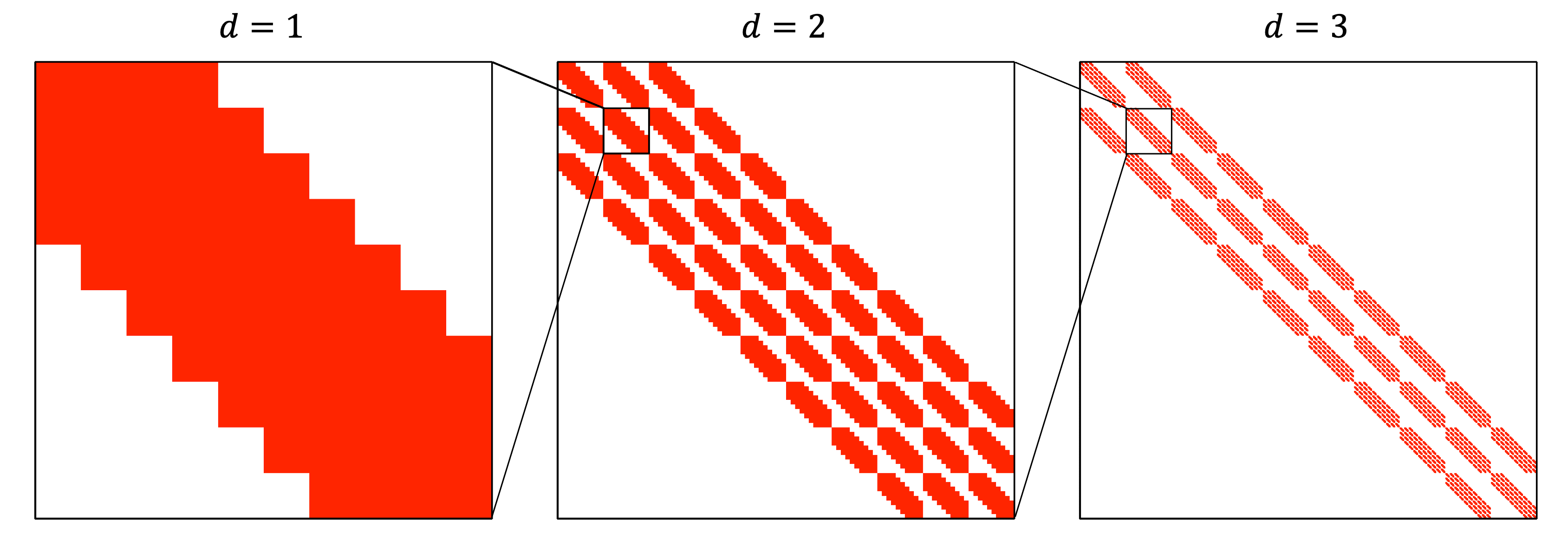}
    \caption{Illustrations of the preserved $\k$-zones $\calH_d(k)$ for univariate lattice $\calS_1(10)$ ($d=1$ and $k=10$) (left), the $2$-order lattice $\calS_2((10,10))$ (middle); and the $3$-order lattice $\calS_3((10,10,10))$ (right). The entries that are filled with red or white represent $\bfdelta_{ij}\in\calH_d(\k)$ or $\bfdelta_{ij}\notin\calH_d(\k)$, respectively.}
    \label{fig:hyper-rectangle}
\end{figure}

Figure \ref{fig:hyper-rectangle} illustrates the preserved $\k$-zones $\calH_d(\k)$ for $d=1$, $2$ and $3$, respectively. The figure shows that the preserved $\k$-zones capture the structural dependencies inherent in the tensor data more effectively than existing bandable classes, especially for $d \ge 2$. Specifically, as shown in the inlets of the figure, the preserved $\k$-zone for the $2$-order lattice includes the bandable structures for $d=1$ in its banded area, while the preserved $\k$-zone for the $3$-order lattice includes the bandable-in-bandable structure for $d=2$ in its banded area. It is evidence that a higher order $k$-zone offers richer local dependence structure than a lower order preserved $k$-zone.

\begin{figure}[ht!]
    \centering
    \begin{subfigure}[b]{0.45\textwidth}
        \caption{$\tau(\k)=k_1^{-0.5}+k_2^{-0.3}$}
        \centering
        \includegraphics[width=0.7\textwidth]{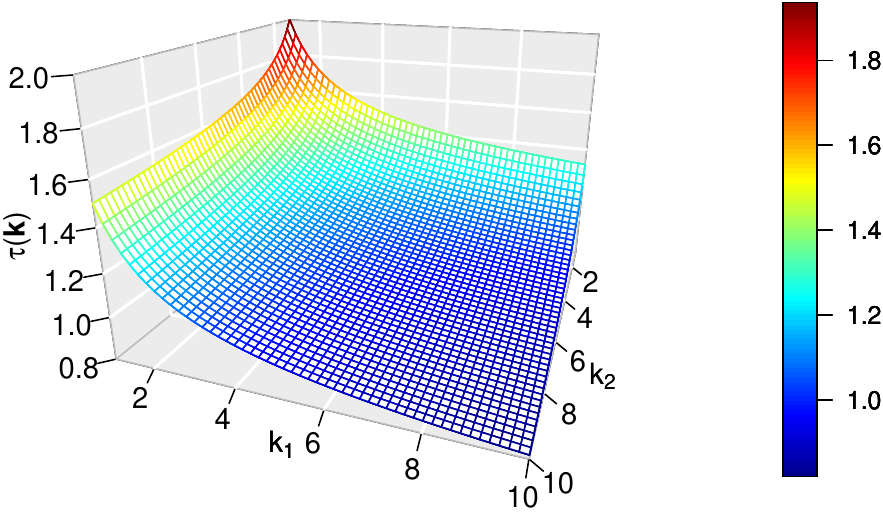}
    \end{subfigure}
    \begin{subfigure}[b]{0.45\textwidth}
        \caption{$\tau(\k)=1.1^{-k_1}+1.2^{-k_2}$}
        \centering
        \includegraphics[width=0.7\textwidth]{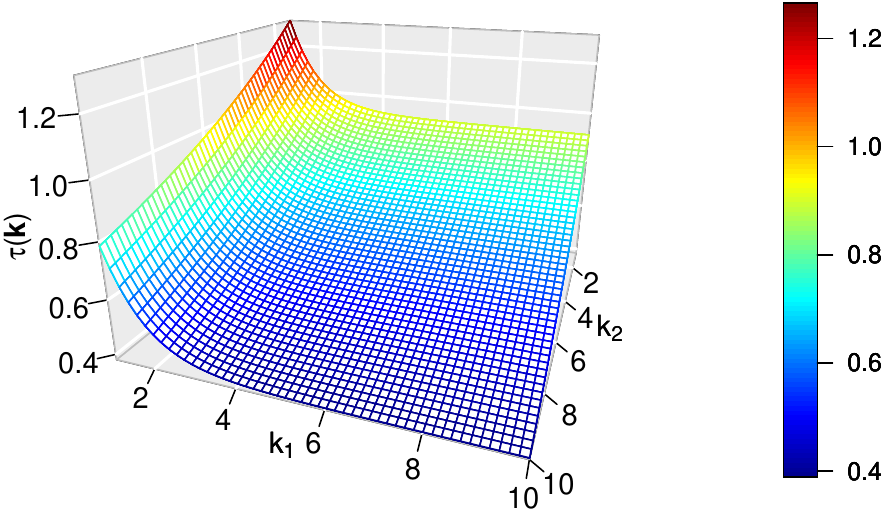}
    \end{subfigure}
    \caption{The covariance decay function $\tau(\k)$ for $d=2$ with $\tau(\k)=k_1^{-0.5}+k_2^{-0.3}$ (a) and $\tau(\k)=1.1^{-k_1}+1.2^{-k_2}$ (b).
    }
    \label{fig:tau}    
\end{figure}

We consider a more general form of the upper bound than $Ck^{-\alpha}$ used in Condition (i) of the bandable covariance class $\calU_1(\alpha,\epsilon,C)$ in \eqref{eq:bandable-class-BL}. Specifically, we define a general $d$-variate \textit{covariance decay function} $\tau(\k)$, which offers patterns of covariance decay. The following assumption regulates the covariance decay function.
\begin{assumption}\label{assume:cov-decay}
    The covariance decay function $\tau(\k)$ is a $d$-variate function defined on $\calH_d(\p)$ and is non-negative, uniformly bounded, marginally non-increasing with respect to each dimension of the lattice and satisfies $\tau(\k)\to0$ if and only if $\min_\ell k_\ell\to\infty$.
\end{assumption}
Assumption \ref{assume:cov-decay} is a regularity condition on the non-increasing covariance of the two grids as they are gradually far away. Figure \ref{fig:tau} displays two examples of the covariance decay functions $\tau(\k)$ for $d=2$, including the polynomial decay $\tau(\k)=\sum_{\ell=1}^dk_\ell^{-\alpha_\ell}\bbI\{0<k_\ell<p_\ell\}$ for $\{\alpha_{\ell}\}_{\ell=1}^d$ being positive numbers and the exponential decay $\tau(\k)=\sum_{\ell=1}^d\beta_\ell^{-k_\ell}\bbI\{k_\ell<p_\ell\}$ for $\{\beta_{\ell}\}_{\ell=1}^d$ larger than $1$. They illustrate the manner in which the sum of covariances diminishes for all grid pairs that are spatially separated by a two-order rectangular region $\calH_2\big((k_1,k_2)\transpose\big)$. The covariance decay function $\tau(\k)$ allows discontinuity at the boundary $k_\ell=p_\ell$ for an $\ell$. For example, for the bandable covariance class $\calU_1(\alpha,\epsilon,C)$, $\tau(k)=Ck^{-\alpha}\bbI(0<k<p)$ which yields $\tau(p)=0$. For the above two examples, the term $k_\ell<p_\ell$ in $\tau(\k)$ prescribes the fact that the $\ell$th dimension no longer contributes to the covariance decay form when $k_\ell$ exceeds $p_\ell$. See the discussion for the general case in Section \ref{sec:optimal-scale} of the SM.

With the above preparation, we propose a covariance class for the tensor data sampled from a $d$-order lattice. Specifically, given a covariance decay function $\tau(\k)$, we define a $d$-order \textit{multi-bandable covariance class}
\begin{eqnarray}
    \label{eq:bandable-class-prop}
    \calU(d,\tau,\epsilon)=\Big\{\bfSigma=[\sigma_{ij}]_{p\times p}: &\text{(i)}& \max_j\sum_{\{i:\bfdelta_{ij}\notin\calH_d(\k)\}}|\sigma_{ij}|\le \tau(\k) \hbox{ for all } \k\in\calH_d(\p),\\
    &\text{(ii)}& \add{0}\le \lambda_{\min}(\bfSigma)\le\lambda_{\max}(\bfSigma)\le \epsilon^{-1}\Big\}, \nn
\end{eqnarray}
for some positive constant $\epsilon$.

The proposed covariance class imposes regularization on the covariance through the covariance structure captured by the preserved $\k$-zone $\calH_d(\k)$. It is noted that $\calU(1,\tau,\epsilon)$ contains the bandable covariance class $\calU_1(\alpha,\epsilon,C)$ in \eqref{eq:bandable-class-BL}, with the polynomially decayed $Ck^{-\alpha}$ in Condition (i) replaced with a general covariance decay function $\tau(\k)$.  In the meanwhile, the separably bandable covariance classes $\calU_{3,1}(\alpha_1,\alpha_2,\epsilon,C)$ in \eqref{eq:bandable-class-ZSK} is a special case of $\calU(2,\tau,\epsilon)$, as it prescribed the polynomial decayed only and assumes additional separability in the two directions of the bivariate lattice.
In contrast, the proposed covariance class for $d=2$ permits general covariance decay patterns and prescribes the bandable-in-bandable covariance structure in the matrix data without the separability assumption.

\subsection{Localization Estimator}
\label{sec:local-est}

Recall that past studies have proposed the banding and tapering estimators for the bandable covariance class $\calU_1(\alpha,\epsilon,C)$ and $\calU_2(\alpha,\epsilon,C)$, and the separably banding or tapering estimator for the separably bandable covariance class $\calU_{3,q}(\alpha_1,\alpha_2,\epsilon,C)$. To better address the complex covariance structure induced by the proposed multi-bandable covariance class $\calU(d,\tau,\epsilon)$, a general \hd\ covariance estimator will be developed, which is the main focus of this subsection.

Let $\S_n=[\hat{\sigma}_{ij}]_{p\times p}$ be the sample covariance and denote by $\a/\b$ the element-wise division of two vectors $\a$ and $\b$. We propose a \textit{localization} estimator for the multi-bandable covariance class $\calU(d,\tau,\epsilon)$
\begin{eqnarray}\label{eq:localization-estimator}
    \calL_h(\S_{n};\k_h)=\big[\hat{\sigma}_{ij}h(\bfdelta_{ij}/\k_h)\big]_{p\times p},
\end{eqnarray}
where $h(\bfdelta_{ij}/\k_h)$ is a weight taking value in $[0,1]$ for the $(i,j)$-entry and is determined by the absolute coordinate difference $\bfdelta_{ij}$, a vector of scaling parameters $\k_h=(k_{h1},k_{h2},\dots,k_{hd})\transpose$ and a $d$-variate \textit{localization function} $h$.

The estimator $\calL_h(\S_n;\k_h)$ is motivated by the localization technique in the data assimilation area, where to cope with the spurious correlation, only the covariance in a local area of each assimilated grid is preserved \citep{Houtekamer2001, Hamill2001, Furrer2007}. Drawing on this, the proposed estimator addresses regularization on the underlying covariance structure through two ingredients, the localization function $h$ and a scaling vector $\k_h$. 

The localization function $h$ is usually empirically designated with a goal to account for the diminishing covariance of two grids as they are gradually separated. The following assumption is on the localization function $h$.
\begin{assumption}\label{assume:cov-local}
    (i) For $\z=(z_1,z_2,\dots,z_d)$ with $z_\ell\ge 0$ for $\ell=1,\dots,d$, the $d$-variate localization function $h(\z)$ satisfies $h(\0_d)=1$, $h(\z)=0$ if $z_{\ell}\ge1$ for an $\ell\in\{1,2,\dots,d\}$, and $h(\z)$ is marginally non-increasing with respect to each $z_{\ell}$.
    (ii) There exist constants $c_1,c_2,\dots,c_d\in(0,1)$ such that $h(\z)=1$ for $0\le z_\ell\le c_\ell$ and all $\ell=1,\dots,d$.
\end{assumption}
Assumption \ref{assume:cov-local} (i) is similar to Assumption \ref{assume:cov-decay} for the covariance decay function $\tau(\k)$, which prescribes that $h$ should be non-negative, compactly supported on $[0,1]^d$ and marginally non-increasing. Part (ii) is to control the bias of the estimation when using the localization function $h$, which demonstrates that $\hat{\sigma}_{ij}$ would be fully preserved when the standardized distance between the $i$th and the $j$th grids in the $\ell$th direction is no larger than $c_\ell$. 

To account for the variation of the estimation error under the spectral norm, we require another restriction on the variation of the localization function $h$, which requires the notion of Vitali variation. The Vitali variation \citep{Vitali1908} is a generalization of the total variation to multi-dimensional spaces. Specifically, given two distinct points $\a=(a_{1},a_{2},\dots,a_{d})\transpose$ and $\b=(b_{1},b_{2},\dots,b_{d})\transpose$ that satisfies $a_{\ell}\le b_{\ell}$ for $\ell=1,\dots,d$, suppose $[\a,\b]=[a_1,b_1]\times[a_2,b_2]\times\dots\times[a_d,b_d]$, we denote the quasi-volume
\begin{equation}
    \label{eq:quasi-volume}
    \text{qVol}(h;[\a,\b])=\sum_{j_1=0}^{J_1}\dots\sum_{j_d=0}^{J_d}(-1)^{j_1+\dots+j_d}h\big((b_{1}+j_1(a_1-b_1),,\dots,b_{d}+j_d(a_d-b_d))\transpose\big)
\end{equation}
where $J_\ell=\bbI\{a_\ell\neq b_\ell\}$ for each $\ell=1,\dots,d$. Then, given $d$ univariate partitions $0=a_{\ell,0}<a_{\ell,1}<\dots<a_{\ell,N_\ell}=1$ for $\ell=1,\dots,d$ and some positive integers $N_\ell$, let $\calP$ be a collection of all sets of the form $\calA=A_1\times A_2\times\dots\times A_d$ where $A_\ell=[a_{\ell,n_\ell},a_{\ell,n_\ell+1}]$ for each $\ell=1,\dots,d$ and $0\le n_\ell\le N_\ell-1$, the Vitali variation of $h(\z)$ is defined as
\begin{equation}
    \label{eq:vitali-variation}
    \text{ViV}(h):=\sup_{\calP}\sum_{\calA\in\calP}|\text{qVol}(h;\calA)|.
\end{equation}
Particularly, if $h$ is $d$-times continuously differentiable on $[0,1]^d$, namely, the mixed partial derivative $\partial^d h/(\partial z_1 \partial z_2 \cdots \partial z_d)$ exists and is continuous, then 
\begin{equation*}
    \text{ViV}(h)=\int_{0}^1\int_{0}^1\dots\int_{0}^1\Big|\frac{\partial^d h(\z)}{\partial z_1\partial z_2\dots \partial z_d}\Big|\mathrm{d}z_1\mathrm{d}z_2\dots \mathrm{d}z_d.
\end{equation*}

\begin{assumption}\label{assume:cov-local-2}
    The localization function $h$ satisfies $\text{ViV}(h)<C$ for a constant $C>0$.
\end{assumption}

Assumption \ref{assume:cov-local-2} is valid for a wide range of functions. Two sufficient conditions for the bounded Vitali variation are as below. One is that $h(\z)$ is continuous in $[0,1)^d$ except for some isolated discontinuous points and there exists an axis-aligned rectangular partition of $[0,1]^d$ denoted as $\{\calD_b\}$ such that $h(\z)$ is smooth in each $\calD_b$ and satisfies $\int_{\z\in \calD_b}\big|\frac{\partial^d h(\z)}{\partial \z_1\z_2\dots \partial \z_d}\big|\mathrm{d}\z<C$ for some constant $C$. The other sufficient condition is that $h(\z)$ is a multiplicative function $\prod_{\ell=1}^dh_\ell(z_\ell)$ with the total variation of each $h_\ell(z_\ell)$ being finite. One may refer to \cite{Fang2021} for detailed discussions.

For $d=1$, the banding function \eqref{eq:banding-estimator} and the tapering function \eqref{eq:linear-taper-estimator} are two special cases of the localization function $h$. In the meanwhile, the localization functions $h(z)$ for $d=1$ can also permit non-linearity in $z\in(c,1)$ for some constant $0<c<1$. The localization function $h$ naturally extends the banding and the tapering function by adapting to the tensor data in a multi-order lattice through the absolute coordinate difference $\bfdelta_{ij}$. For example, that $\tilde{\bfSigma}(k_1,k_2)$ on the right-hand side of the separably banding or tapering estimator \eqref{eq:ZSK-approximation} employed a doubly banding function $\prod_{\ell=1}^2\bbI\{z_\ell<1\}$ and a doubly tapering function $\prod_{\ell=1}^2\varphi(z_\ell;0.5,1)$ for $d=2$, respectively. One can see that Assumptions \ref{assume:cov-local} and \ref{assume:cov-local-2} hold for both functions. On the other hand, the two functions are specific examples of a class of multiplicative localization functions $h(\z)=\prod_{\ell=1}^dh_\ell(z_\ell)$ with each $h_\ell(z_\ell)$ satisfying Assumptions \ref{assume:cov-local} and \ref{assume:cov-local-2}. The merit of such a design is that the heterogeneity among different dimensions of the lattice can be addressed by different covariance decay patterns offered by $h_\ell$ and $k_{\ell}$, respectively. 

Specifically, we define a \textit{multi-banding} estimator associated with a set of banding widths or scaling vector $\k=(k_1,k_2,\dots,k_d)\transpose$ as
\begin{equation}
    \label{eq:multi-banding-estimator}
    \hat\bfSigma_\k:=\Big[\hat\sigma_{ij}\bbI\big\{\delta_{ij\ell}<k_\ell \hbox{ for all }\ell=1,\dots,d\big\}\Big]_{p\times p}. 
\end{equation}
The multi-banding estimator is a specialized localization estimator where the localization function $h(\z)=\prod_{\ell=1}^d\bbI\{z_\ell<1\}$ and the scaling vector $\k$, which is analogous to $\k_h$ in the general localization estimators \eqref{eq:localization-estimator}. 

In the data assimilation area, a commonly used localization function $h$ is the Gaspari-Cohn (GC) function \citep{Gaspari1998}
\begin{eqnarray}
    \label{eq:GC-func}
    \text{GC}(z)=\left\{\begin{array}{lll}
         1-\frac{5}{3}z^2+\frac{5}{8}z^3+\frac{1}{2}z^4-\frac{1}{4}z^5,  &\,0\le z\le 1; \\
        -\frac{2}{3}{z}^{-1}+ 4-5z+\frac{5}{3}z^2+\frac{5}{8}z^3-\frac{1}{2}z^4+\frac{1}{12}z^5, &\,1<z\le 2; \\
        0 , &\, z\ge 2. 
    \end{array}\right.
\end{eqnarray}
In practice, the GC function can impose a weight $\text{GC}(2\|\bfdelta_{ij}/\k_{\text{GC}}\|)$ on the $(i,j)$th entries of the sample covariance $\S_n$, which decays with respect to the $L_2$ distance and is homogeneous among the $d$ directions of the lattice after scaling by a scaling vector $\k_{\text{GC}}$. Although the GC function does not satisfy Assumption \ref{assume:cov-local} (ii), the bias of the localization estimator using the GC function can be negligible under mild conditions, since the weights for the covariance between two relatively close grids are quite close to $1$.

\begin{figure}[ht!]
    \centering
    \begin{subfigure}[b]{0.48\textwidth}
        \caption{multiplicative banding function ($d=2$)}
        \centering
        \includegraphics[height=3cm]{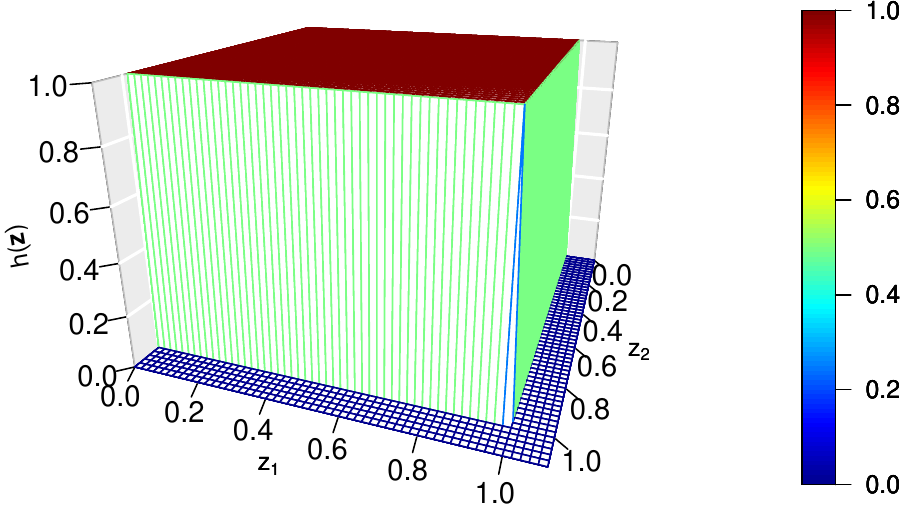}
    \end{subfigure}
    \begin{subfigure}[b]{0.48\textwidth}
        \caption{multiplicative tapering function ($d=2$)}
        \centering
        \includegraphics[height=3cm]{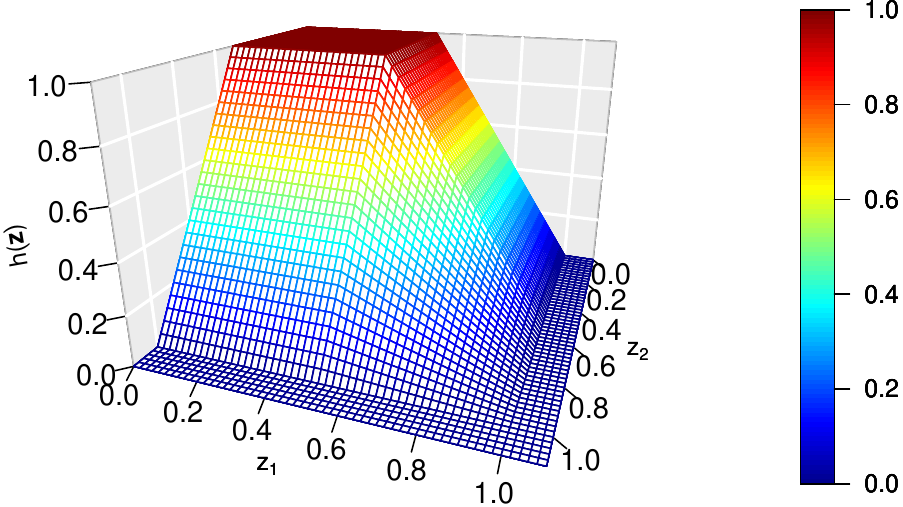}
    \end{subfigure}
    \begin{subfigure}[b]{0.48\textwidth}
        \caption{GC function ($d=1$)}
        \centering
        \includegraphics[height=3cm]{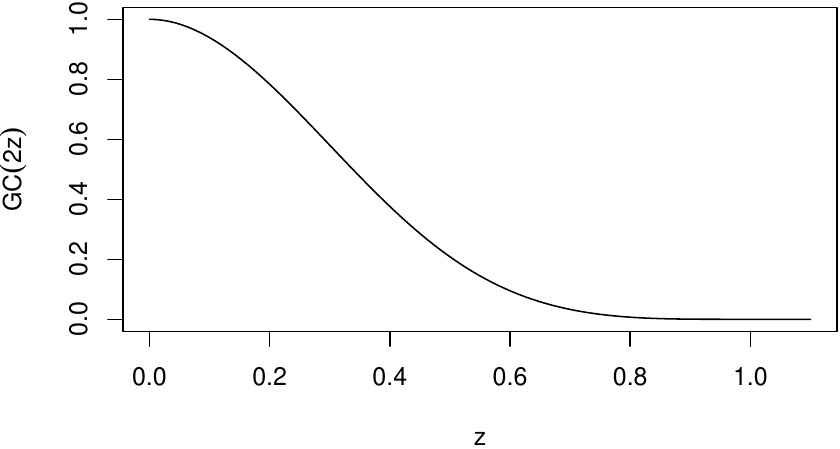}
    \end{subfigure}
    \begin{subfigure}[b]{0.48\textwidth}
        \caption{GC function ($d=2$)}
        \centering
        \includegraphics[height=3cm]{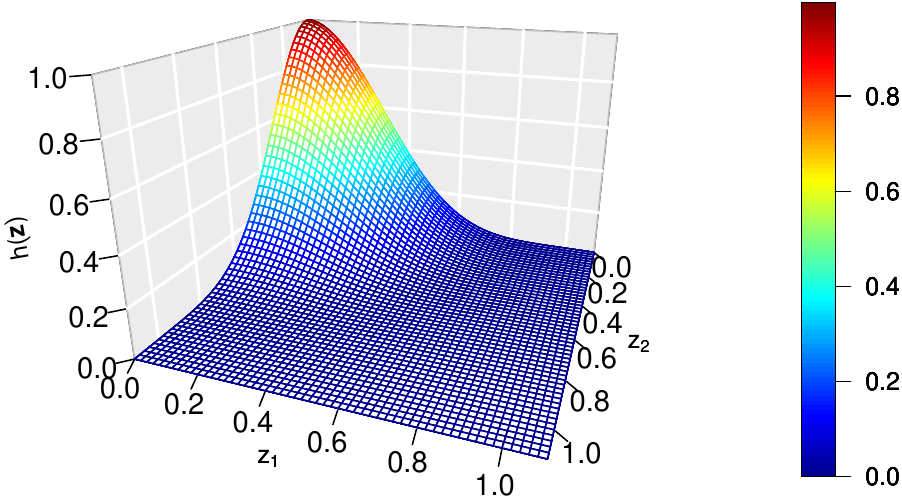}
    \end{subfigure}    
    \caption{Four examples of the localization function $h(\z)$ with $h$ being the multiplicative banding function $\prod_{\ell=1}^2\bbI\{z_\ell<1\}$ (a), the multiplicative tapering function $\prod_{\ell=1}^2\varphi(z_\ell;0.5,1)$ (b), the GC function $\text{GC}(2z)$ for $d=1$ (c) and $\text{GC}(2\sqrt{z_1^2+z_2^2})$ for $d=2$ (d).}
    \label{fig:local-func}    
\end{figure}

Figure \ref{fig:local-func} displays some of the above-mentioned localization functions, including the multiplicative banding and tapering functions for $d=2$, and the GC function for $d=1$ and $2$. One can see that different choices of the localization functions can offer different regularization weights with respect to the $2$-order absolute coordinate $\bfdelta_{ij}$ after scaling by $\k_h$.

The scaling vector $\k_h$ determines the maximum allowable separation distance in each coordinate direction beyond which the covariance between two grids is set to zero. A choice of $\k_h$ is based on an cross-validation measure on the covariance estimation with respect to the banding width or scaling vectors via the data splitting as in \cite{Bickel2008} and \cite{Zhang2023}, whose detail will be outlined in Section \ref{sec:simulation}. For $d=1$ and $h$ being the banding or tapering function, \cite{Qiu2015} provided a more vigorous selection scheme by minimizing a standardized expected square of the Frobenius loss in the estimation of the covariance matrix.

It is noted that there is no guarantee that the proposed localization estimator \eqref{eq:localization-estimator} is positive-semidefinite. One may avoid this by reconstructing the localization estimator $\calL_h(\S_{n};\k_h)$ via eigenvalue-decomposition with the negative eigenvalues being trimmed off.

\section{Theoretical Results}
\label{sec:main-result}
We establish the consistency of the localization estimator under the spectral and the Frobenius norms, which requires the following assumption.
\begin{assumption}
    \label{assume:sub-Gaussian} 
    The vectorized data $\X_1$ follows a sub-Gaussian distribution such that  
    \begin{eqnarray}\label{eq:sub-Gaussian}
        \mathbb{P}\big\{\big|\mathbf{v}\transpose\big(\X_1-\bbE \X_1\big)\big|>t\big\}\le\exp(-\rho t^2/2)\hbox{ for all } t>0 \hbox{ and } \|\mathbf{v}\|=1,
    \end{eqnarray}
    for some constant $\rho>0$.
\end{assumption}

We first investigate the consistency of the localization estimator under the spectral norm. Specifically, we define $V(\k)=\prod_{\ell=1}^dk_\ell$ for a scaling vector $\k$ and denote by $\calI_d(\k)$ a collection of all the positive integer points in $\calH_d(\k)$. Let $C$ be a general positive constant that may vary in different contexts. Then, the following lemma represents the localization estimator $\calL_h(\S_n;\k_h)$  by a set of the multi-banding estimators $\{\hat\bfSigma_\k\}_\k$.

\begin{lemma}\label{lemma:upper-spectral-general}
    Under Assumptions \ref{assume:cov-local} and \ref{assume:cov-local-2}, the localization estimator can be written as
    \begin{equation}\label{eq:band-estimator-splitting}
        \calL_h(\S_n;\k_h)=\sum_{\k\in\calI_d(\k_h)}w(\k)\hat\bfSigma_\k,
    \end{equation}
    where $\{\hat\bfSigma_\k\}_\k $
    are the multi-banding estimator defined in \eqref{eq:multi-banding-estimator} and $\{w(\k)\}_\k$ are weights which satisfy
    \begin{eqnarray}
        \label{eq:match-weight}
        w(\k): &=&\text{qVol}(h;[(\k-\1_d)/\k_h),\k/\k_h] \\
        &=&\sum_{\u=(u_1,u_2,\dots,u_d)\in\{0,1\}^d}(-1)^{u_1+u_2+\dots+u_d}h\{(\k-\u)/\k_h\}. \nn
    \end{eqnarray}
\end{lemma}
Lemma \ref{lemma:upper-spectral-general} demonstrates that any localization estimator can be decomposed into a weighted sum of a larger number of multi-banding estimators $\{\hat\bfSigma_\k\}_\k$, with the weights $\{\w(\k)\}_\k$ corresponding to the quasi-volume defined in \eqref{eq:quasi-volume}. The decomposition will serve as a construction to facilitate theoretical analysis, for instance that reported in the following lemma.

\begin{lemma}\label{lemma:upper-spectral-banding}
    Under Assumptions \ref{assume:cov-decay} and \ref{assume:sub-Gaussian}, for $\log p=o(n)$ and $V(\k)=o(n)$, the multi-banding estimator in \eqref{eq:multi-banding-estimator} satisfies
    \begin{equation}
        \label{eq:band-variance}
        \sup_{\bfSigma\in\calU(d,\tau,\epsilon)}\bbE\|\hat\bfSigma_\k-\bbE\hat\bfSigma_\k\|^2\le C\frac{\log p+V(\k)}{n}.
    \end{equation}
\end{lemma}
Lemma \ref{lemma:upper-spectral-banding} establishes that $\bbE\|\hat\bfSigma_\k-\bbE\hat\bfSigma_\k\|^2$, which is a kind of variation of the multi-banding estimator under the spectral norm, is controlled by $\log p/n$ and $V(\k)/n$. We note in passing that, for $d=1$, the variation of the banding estimator $\calB_k(\S_n)$ under the spectral norm achieves an upper bound of the {variation} as $(\log p+k)/n$, which improves the results of $k\log p/n$ in \cite{Bickel2008}. Compared with the analysis in \cite{Cai2010}, they characterizes the variation of the tapering estimator $\calT_{k}(\S_n)$ through their Lemmas 1 and 2, which decomposes $\calT_{k}(\S_n)$ into an average of matrices that are sum of disjoint block matrices and derives the upper bound of each block's variance. The proof, when applied to the banding estimator, would introduce an additional factor of $k$ in the {variation's upper bound}, that prevented attaining the optimal rate of convergence $n^{-(2\alpha)/(2\alpha+1)}$ in \eqref{eq:convergence-rate-CZZ} when $p>n^{1/(2\alpha+1)}$ for the banding estimator. We develop an alternative proof strategy by partitioning the multi-banding estimator into multiple covariance or cross-covariance matrices of size $V(\k)\times V(\k)$, with the recently developed random matrix theory in \cite{Park2021} being applied to analyze the variation under the spectral norm.

The following theorem provides the consistency of the localization estimator under the spectral norm.
\begin{theorem}
    \label{thm:minimax-spectral-upper}
    Under Assumptions \ref{assume:cov-decay}, \ref{assume:cov-local}, \ref{assume:cov-local-2} and \ref{assume:sub-Gaussian}, for $\log p=o(n)$ and $V(\k_h)=o(n)$, the localization estimator \eqref{eq:localization-estimator} satisfies
    \begin{equation}
        \label{eq:minimax-spectral-upper-1}
        \sup_{\bfSigma\in\calU(d,\tau,\epsilon)}\bbE\big\|\calL_h(\S_n;\k_h)-\bfSigma\big\|^2 \le C\tau^2(\k_h\circ\c)+C\frac{\log p+V(\k_h)}{n},
    \end{equation}
    where $\c=(c_1,c_2,\dots,c_d)\transpose$ is defined in Assumption \ref{assume:cov-local} (ii). Specifically, the localization estimator with the scaling vector $\k_h = \arg\min_{\k\in\calH_d(\p)}\allowbreak\{\tau^2(\k)+V(\k)/n\}$ satisfies
    \begin{equation}
        \label{eq:minimax-spectral-upper-2}
        \sup_{\bfSigma\in\calU(d,\tau,\epsilon)}\bbE\big\|\calL_h(\S_n;\k_h)-\bfSigma\big\|^2 \le C\varepsilon_{n,\p}+C\frac{\log p}{n},
    \end{equation}
    where $\varepsilon_{n,\p}=\min_{\k\in\calH_d(\p)} \{\tau^2(\k)+V(\k)/n\}$.
\end{theorem}

Theorem \ref{thm:minimax-spectral-upper} demonstrates that consistency of the localization estimator \eqref{eq:localization-estimator} to the covariance matrix $\bfSigma$ if the dimension of the grids satisfies $\log p=o(n)$ and the scaling vector satisfies $V(\k_h)=o(n)$. Specifically, the estimation error of the localization estimator in \eqref{eq:minimax-spectral-upper-1} is contributed by two terms: $\tau^2(\k_h\circ\c)$ and $n^{-1}\{\log p+V(\k_h)\}$. The $\tau^2(\k_h\circ\c)$ term represents the bias introduced by the entries $\{\hat{\sigma}_{ij}h(\bfdelta_{ij}/\k_h)\}$ in $\calL_h(\S_n;\k_h)$ due to the weight $h(\bfdelta_{ij}/\k_h)$ being less than $1$, which converges to $0$ according to Assumption \ref{assume:cov-decay} as long as the scaling parameters $k_{h\ell}\to\infty$ as the sample size $n$ and dimensions $p_\ell\to\infty$. In the meanwhile, the latter term illustrates a form of variation under the spectral norm.

Theorem \ref{thm:minimax-spectral-upper} imposes restrictions on the localization functions only through Assumptions \ref{assume:cov-local} and \ref{assume:cov-local-2}. Therefore, the aforementioned examples of the localization function in Section \ref{sec:local-est} are all able to achieve the rate of convergence \eqref{eq:minimax-spectral-upper-1} in Theorem \ref{thm:minimax-spectral-upper}, including the case of $d=1$ that covers the univariate banding and the tapering functions as well as the multiplicative banding or tapering functions for $d\ge 2$. On the other hand, non-linearity is also permitted for the localization functions as prescribed in Assumption \ref{assume:cov-local}.

The first term of the upper bound $\varepsilon_{n,\p}$ in \eqref{eq:minimax-spectral-upper-2} provides a general form of the ``bias-variance" trade-off between the covariance decay function $\tau(\k)$ and $V(\k)/n$, the ``volume" of the preserved $\k$-zone for the preserved entries divided by the sample size. Hence, $\varepsilon_{n,\p}$ can be guaranteed to converge to $0$ under Assumption \ref{assume:cov-decay} as the dimensions $\{p_\ell\}$ and the sample size $n$ increase. Specifically, the upper bound $\varepsilon_{n,\p}$ can be attained by properly choosing the localization scaling vector $\k_h$ for the localization function $h$, which is determined by solving a bounded optimization problem. The two subscripts $n$ and $\p$ to $\varepsilon_{n,\p}$  demonstrate that the error bound relies on both sample size $n$ and dimensions $\p$. Specifically, if $n=o(p_\ell)$ for $\ell=1,\dots,d$, $\varepsilon_{n,\p}$ is only relevant to the sample size $n$.

When $d=1$ and $\tau(k)=Ck^{-\alpha}\bbI\{0<k<p\}$,
\begin{equation}
    \varepsilon_{n,p}\asymp\min_{0\le k\le p}\big(k^{-2\alpha}\bbI\{0<k<p\}+k/n\big)\asymp\min\{n^{-\frac{2\alpha}{2\alpha+1}},p/n\} \label{eq:univariate1}
\end{equation}
by choosing $k_h\asymp\min\{n^{1/(2\alpha+1)},p\}$. The result generalizes the convergence rate \eqref{eq:convergence-rate-CZZ} in \cite{Cai2010} to the localization estimators equipped with general localization functions that satisfy Assumption \ref{assume:cov-local} and \ref{assume:cov-local-2}, which covers the banding estimator $\calB_k(\S_n)$ in \cite{Bickel2008} as a special case. Combined with the minimax lower bound established in \cite{Cai2010}, this implies that the rate given in (\ref{eq:univariate1}) is actually the minimax optimal rate for the banding estimator under the spectral norm.

\begin{figure}
    \centering
    \includegraphics[width = 0.5\textwidth]{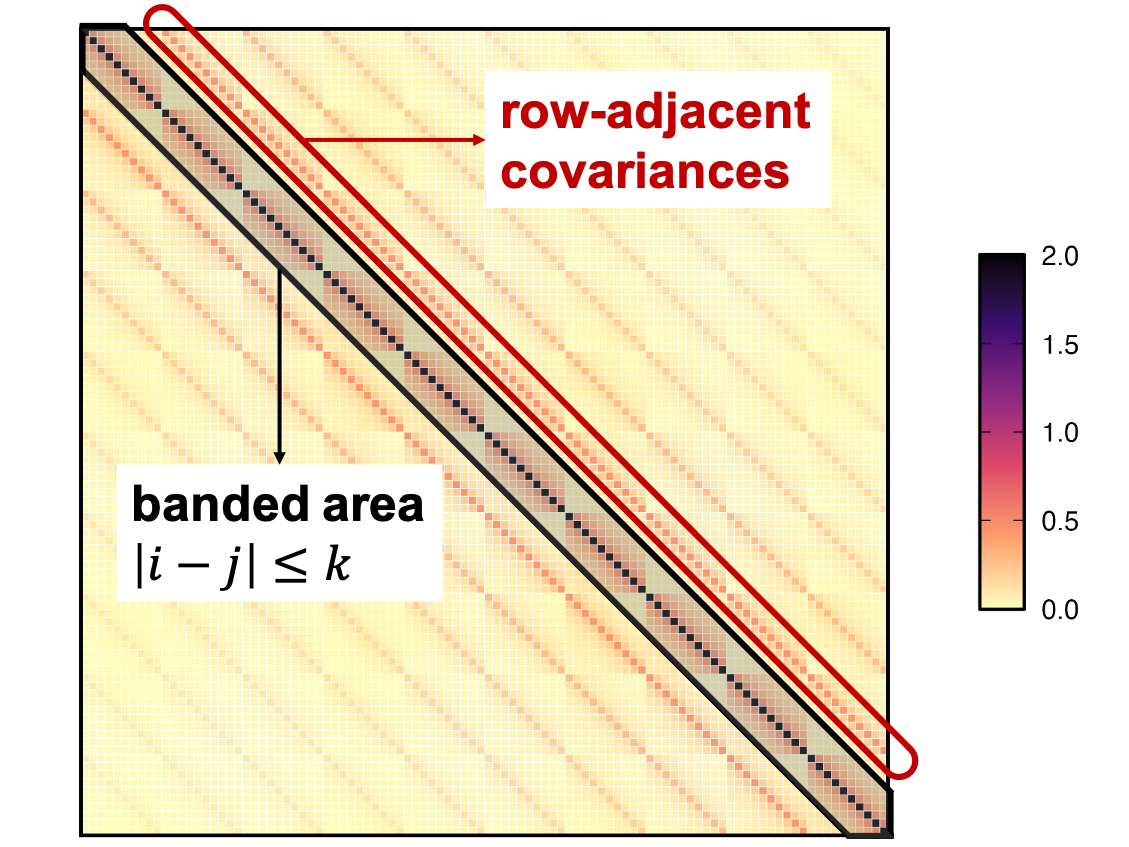}
    \caption{Heatmap of $\bfSigma = (\sigma_{ij})_{p\times p}\in\calU(2,\tau,\epsilon)$ for a $10\times 10$ matrix data vectorized in the column-major order, with the entries satisfies $\sigma_{ij}=\prod_{\ell=1}^2(1+\delta_{ij\ell})^{-\alpha_\ell-1}$ for $\alpha_1=0.1$ and $\alpha_2=0.2$. The gray region is a banded area  $|i-j|\le k$ for the banding estimator with a banding width $k<p_1$, and the red colored bands illustrate the covariances between adjacent grids along rows, which decay rather slowly that prevents the bandable condition in (2.3). }
    \label{fig:2D-banding-example}
\end{figure}

Another finding is that the banding estimator $\calB_k(\S_n)$ of \cite{Bickel2008}, which are effective for $d=1$, may not guarantee consistency for $d\ge 2$. We illustrate this issue for an example of $d = 2$, where the variables are observed on the lattice $\calS_2(\p)=\{1,\dots,p_1\}\times\{1,\dots,p_2\}$. Let $\bfSigma = (\sigma_{ij})$ denote the covariance matrix of the variables, vectorized in the column-major order of $\calS_2(\p)$. Suppose $\sigma_{ij}=\prod_{\ell=1}^2(1+\delta_{ij\ell})^{-\alpha_\ell-1}$ for $\delta_{ij\ell} = 0, 1, \dots, p_\ell-1$, $\ell=1$ and $2$, where $\alpha_1$ and $\alpha_2$ are positive constants. The heatmap of this covariance matrix is shown in Figure \ref{fig:2D-banding-example}. It can be verified that this $\bfSigma$ satisfies the proposed multi-bandable covariance class in (\ref{eq:bandable-class-prop}) such that $\bfSigma\in\calU(2,\tau,\epsilon)$ with $\tau(\k)=C\sum_{\ell=1}^2(1+k_\ell)^{-\alpha_\ell}$. However, this covariance does not satisfy the bandable condition required in \eqref{eq:bandable-class-BL}, since
\begin{equation}    
    \label{eq:cov-example}
    \sum_{|i - j| \geq p_1} |\sigma_{ij}| \geq |\sigma_{i\, i + p_1}| \not\to 0 \mbox{ \ for each given $i$,}
\end{equation}
as $p_1 \to \infty$. The reason for $|\sigma_{i\, i + p_1}| \not\to 0$ is because that $\sigma_{i\, i + p_1}$ is the covariance between the variables at the locations $(s_1, s_2)\transpose$ and $(s_1, s_2 + 1)\transpose$, which are adjacent in a row of $\calS_2(\p)$. This shows that the multi-bandable class in (\ref{eq:bandable-class-prop}) for tensor data may not satisfy the bandable class in \eqref{eq:bandable-class-BL}, which implies that there is no guarantee for consistent estimation of $\bfSigma$ by the banding estimator $\calB_k(\S_n)$ of \cite{Bickel2008}.  

In fact, to make the mean squared error (MSE) of $\calB_k(\S_n)$ converge to $0$, it is required that both the bias term $\|\calB_k(\bfSigma) - \bfSigma\|$ and the variance term $\mathbb{E}\|\calB_k(\S_n)-\bbE\calB_k(\S_n)\|^2$ converging to $0$ as $n, p_1, p_2 \to \infty$. Specifically for estimating the covariance matrix given above, it can be shown similarly to (4.8) that the bias term satisfies $\|\calB_k(\bfSigma)-\bfSigma\| \ge \max_{i} |\sigma_{i\, i + p_1}| \not\rightarrow 0$ for $k < p_1$, while the variance is bounded by $\mathbb{E}\|\calB_k(\S_n)-\bbE\calB_k(\S_n)\|^2 = O\big( \{k + \log (p_1p_2)\} / n \big)$ from (\ref{eq:upper-spectral-variance}) in the SM. These imply the inconsistency of the banding estimator for estimating the particular $\bfSigma$. Specifically, from the illustration in Figure \ref{fig:2D-banding-example}, the banded area $|i-j|\le k$ with $k<p_1$ would exclude row-adjacent covariances $\{\sigma_{i\, i+p_1}\}$. This omission would lead to a non-ignorable bias of $\calB_k(\S_n)$. However, choosing $k > p_1$ would make the variance term not diminish if $p_1 > n$. Therefore, the MSE of $\calB_k(\S_n)$ would not converge to $0$ if $p_1 > n$.

The first term of the upper bound $\varepsilon_{n,\p}$ in \eqref{eq:minimax-spectral-upper-2} can have an explicit form in the following cases, where the specific details can be found in Section \ref{sec:optimal-scale}. 

\begin{example}
    [Polynomially decayed covariances]
    We consider an additive and polynomially decayed setting with
    \begin{equation}
        \label{eq:multi-poly-cov}
        \tau(\k)=C\sum_{\ell=1}^d\big[k_\ell^{-\alpha_\ell}\bbI\{0< k_\ell<p_\ell\}+\bbI\{k_\ell=0\}\big]
    \end{equation}
    for some positive constants $\{\alpha_j\}_{j=1}^d$. The upper bound term $\varepsilon_{n,\p}$ can be obtained by solving
    \begin{equation}
        \label{eq:varepsilon-hetero-poly}
        \min_{0< k_\ell \le p_\ell}\Big\{\Big(\sum_{\ell=1}^dk_\ell^{-\alpha_\ell}\bbI\{0< k_\ell<p_\ell\}\Big)^2+n^{-1}\prod_{\ell=1}^dk_\ell\Big\},
    \end{equation}
    in which the solution is within the close set $\calH_d(\p)$ and the objective function is discontinuous at the boundary $k_\ell=p_\ell$ for $\ell\in\{1,2,\dots,d\}$. Then, if $p_\ell>n^{\alpha_\ell^{-1}(2+\sum_{\ell=1}^d\alpha_\ell^{-1})^{-1}}$ for $\ell=1,\dots,d$, it follows that $\varepsilon_{n,\p}\asymp n^{-2(2+\sum_{\ell=1}^d\alpha_\ell^{-1})^{-1}}$ by choosing scaling parameters $k_{h\ell}\asymp n^{\alpha_\ell^{-1}(2+\sum_{\ell=1}^d\alpha_\ell^{-1})^{-1}}$. 
    
    Specifically, if $\alpha_1=\alpha_2=\dots=\alpha_d=\alpha$, then $\varepsilon_{n,\p}=n^{-2\alpha/(2\alpha+d)}$ and the upper bound $n^{-2\alpha/(2\alpha+d)}+\log p/n$ would involve $d$, the dimension of the lattice, through the first term. Then, when the dimension $p=\prod_{\ell=1}^dp_\ell$ is relatively small, say $\log p \le n^{d/(2\alpha+d)}$, $\varepsilon_{n,\p}=n^{-2\alpha/(2\alpha+d)}$ becomes the leading term while $p$ has no effect on the upper bound. In contrast, for large $p$ that satisfies $\log p\ge n^{d/(2\alpha+d)}$, the leading term $\log p/n$ will be influenced by both the dimension $p$ and the sample size $n$. The role of the lattice order $d$ is also reflected in the phase transition of the above two scenarios.

    In the general heterogeneous case, where $\alpha_1, \alpha_2,\dots,\alpha_d$ can be different, each $\{\alpha_\ell\}$ will jointly influence $\varepsilon_{n,\p}$ and the order of the optimal scaling vector $\k_h$. Specifically, if $p_\ell\le n^{\alpha_\ell^{-1}(2+\sum_{\ell=1}^d\alpha_\ell^{-1})^{-1}}$ for an $\ell$, which is a degenerate case in the $\ell$th direction, then it is sufficient to take $k_\ell=p_\ell$ and solve other scaling parameters in \eqref{eq:varepsilon-hetero-poly} for $\ell'\neq\ell$. A detailed discussion for $d=2$ is given in Section \ref{sec:eg-degenerate}, in which the dimensions $p_1$ and $p_2$ appear in the upper bound term $\varepsilon_{n,\p}$ in the degenerate cases.
\end{example}

\begin{example}
    [Exponentially decayed covariances]
    We consider an additive and exponentially decayed setting with $\tau(\k)=C\sum_{\ell=1}^d\beta_\ell^{-k_\ell}\bbI\{k_\ell<p_\ell\}$ for some constants $\beta_1,\beta_2,\dots,\beta_d>1$. The upper bound term $\varepsilon_{n,\p}$ can be obtained by solving
    \begin{equation}
        \label{eq:varepsilon-hetero-exp}
        \min_{0< k_\ell \le p_\ell}\Big\{\Big(\sum_{\ell=1}^d\beta_\ell^{-k_\ell}\bbI\{k_\ell<p_\ell\}\Big)^2+n^{-1}\prod_{\ell=1}^dk_\ell\Big\}.
    \end{equation}
    Specifically, if $\beta_1=\beta_2=\dots=\beta_d=\beta$ and $p_\ell>\log n$ for all $\ell$, then minimizing \eqref{eq:varepsilon-hetero-exp} yields to choose $k_{h\ell}\asymp \log n$ for all $\ell=1,\dots,d$, which leads to $\varepsilon_{n,\p}=O\{(\log n)^d/n\}$. In particular, for $d=1$, $\varepsilon_{n,p}=\min\{\log n / n, p/n\}$ by choosing $k_h\asymp\min\{\log n,p\}$. One can see that the exponentially decayed $\tau(\k)$ offers a much faster covariance decay rate than the polynomially decayed case, and thus requires a more restrictive regularization with the localization scaling parameters of a smaller order.
\end{example}

\begin{remark}
    The results in Theorem \ref{thm:minimax-spectral-upper} are also suitable for data from a class of irregular lattices. Denote by $\tilde\calS_d(p')=\{\tilde\s_1,\tilde\s_2,\dots,\tilde\s_{p^\prime}\}$ a $d$-order irregular lattice that is defined as a set of $p^\prime$ randomly distributed and arranged $d$-variate coordinates. Then, if $\tilde\calS_d(p')\subseteq\calS_d(\p)$ with $\p=(p_1,p_2,\dots,p_d)\transpose$. Theorem \ref{thm:minimax-spectral-upper} holds for covariance estimation of tensor data sampled from $\tilde\calS_d(p')$. The irregular lattice is also a common situation in scientific research, for example, in oceanography due to the land and the seabed terrain that may interrupt a regular shape lattice \citep{Moore2011}. The localization estimator \eqref{eq:localization-estimator} can share the same rate of convergence under the spectral norm in \eqref{eq:minimax-spectral-upper-2} at a small cost of rising the dimension $p^\prime$ to $p=\prod_{\ell=1}^dp_\ell$, the number of grids in the smallest regular lattice $\calS_d(\p)$ that contains $\tilde\calS_d(p')$. Specifically, if $\log p=o(n)$, the statistical consistency of the localization estimator for the data from an irregular lattice is guaranteed.
\end{remark}

\add{
% \begin{remark}
%     For the covariance matrix $\bfSigma\in\calU(d,\tau,\epsilon)$ whose minimum eigenvalue is bounded below by a positive constant, the inverse of the localization estimator with the scaling vector $\k_h = \arg\min_{\k\in\calH_d(\p)}\allowbreak\{\tau^2(\k)+V(\k)/n\}$ satisfies
%     \begin{equation*}
%         \bbE\big\|\{\calL_h(\S_n;\k_h)\}^{-1}-\bfSigma^{-1}\big\|^2=O_p\big(\varepsilon_{n,\p}+\log p/n\big).
%     \end{equation*}
%     It then yields that under Assumptions \ref{assume:cov-decay}, \ref{assume:cov-local}, \ref{assume:cov-local-2} and \ref{assume:sub-Gaussian}, for $\log p=o(n)$, the inverse of the localization estimator $\{\calL_h(\S_n;\k_h)\}^{-1}$ is a consistent estimator of $\bfSigma^{-1}$.
% \end{remark}
Another problem of interest is to estimate $\bfSigma^{-1}$. The consistency of the inverse localization estimator can be established for the covariance matrix $\bfSigma\in\calU(d,\tau,\epsilon)$ whose minimum eigenvalue is bounded below by a positive constant. Specifically, let 
\begin{equation*}
    \tilde\calU(d,\tau,\epsilon)=\Big\{\bfSigma: \bfSigma\in\calU(d,\tau,\epsilon) \text{ and } \lambda_{\min}(\bfSigma)>\epsilon\Big\}
\end{equation*}
for some positive constant $\epsilon$. The following proposition provides the consistency of the inverse localization estimator.
\begin{proposition}
    \label{prop:minimax-spectral-inverse}
    Under Assumptions \ref{assume:cov-decay}, \ref{assume:cov-local}, \ref{assume:cov-local-2} and \ref{assume:sub-Gaussian}, for $\log p=o(n)$, the localization estimator with the scaling vector $\k_h = \arg\min_{\k\in\calH_d(\p)}\allowbreak\{\tau^2(\k)+V(\k)/n\}$ satisfies
    \begin{equation*}
        \sup_{\bfSigma\in\tilde\calU(d,\tau,\epsilon)}\bbE\big\|\{\calL_h(\S_n;\k_h)\}^{-1}-\bfSigma^{-1}\big\|^2\le C\varepsilon_{n,\p}+C\frac{\log p}{n}.
    \end{equation*}
\end{proposition}
Proposition \ref{prop:minimax-spectral-inverse} guarantees the consistency of the inverse localization estimator provided that the covariance matrix is well-conditioned within $\tilde\calU(d,\tau,\epsilon)$ and $\log p=o(n)$, with the convergence rate of the inverse estimator matching that of the original localization estimator. Consequently, it extends the applicability of localization methods to the problems requiring precision matrix estimation under a mild sample size condition.
}

Past \hd\ statistics research had paid attention to the optimal rate of convergence under the Frobenius norm, such as in \cite{Cai2010} and \cite{Zhang2023}. In the meanwhile, there were also tuning parameter selection procedures developed based on the expectation of the Frobenius loss \citep{Yi2013, Qiu2015, Sun2024}. 

Our study on the tensor covariance estimation under the Frobenius norm is made for the following covariance class
\begin{eqnarray*}
    \calV(\bfalpha,d,\epsilon,C)=\Big\{ \bfSigma=[\sigma_{ij}]_{p\times p}: && \text{ (i) } |\sigma_{ij}|\le C\prod_{\ell=1}^d \delta_{ij\ell}^{-\alpha_{\ell}-1} \hbox{ for all } \delta_{ij\ell}\neq 0  \\
   && \hbox{ and } \ell=1,\dots,d; \text{ (ii) } \add{0}\le \lambda_{\min}(\bfSigma)\le\lambda_{\max}(\bfSigma)\le \epsilon^{-1}\Big\}, \nn
\end{eqnarray*}
where $\{\alpha_l\}_{l=1}^d$ and $\epsilon$ are positive constants, and $\bfalpha=(\alpha_1,\alpha_2,\dots,\alpha_d)\transpose$. The covariance class $\calV(\bfalpha,d,\epsilon,C)$ replaces Condition (i) of the multi-bandable class $\calU(d,\tau,\epsilon)$ in \eqref{eq:bandable-class-prop} with a more restrictive condition on the off-diagonal entries. One can see that $\calV(\bfalpha,d,\epsilon,C)\subset\calU(d,\tau,\epsilon)$ with the covariance decay function $\tau(\k)=C\sum_{\ell=1}^d\big[k_\ell^{-\alpha_\ell}\bbI\{0< k_\ell<p_\ell\}+\bbI\{k_\ell=0\}\big]$ in \eqref{eq:multi-poly-cov}. The covariance class $\calV(\bfalpha,d,\epsilon,C)$ is inspired by the bandable covariance class $\calU_2(\alpha,\epsilon,C)$ in \cite{Cai2010} and the separably bandable covariance class $\calU_{3,2}(\alpha_1,\alpha_2,\epsilon,C)$ in \cite{Zhang2023}.

The following theorem leads to the consistency of the localization estimator under the Frobenius norm.

\begin{theorem}
    \label{thm:minimax-Frobenius-upper-V}
    Under Assumptions \ref{assume:cov-decay}, \ref{assume:cov-local} and \ref{assume:sub-Gaussian}, if $V(\k_h)=o(n)$, the localization estimator \eqref{eq:localization-estimator} satisfies
    \begin{equation}
        \label{eq:minimax-Frobenius-upper-V-0}
            \sup_{\bfSigma\in\calV(\bfalpha,d,\epsilon,C)}p^{-1}\bbE\big\|\calL_h(\S_n;\k_h)-\bfSigma\big\|_F^2\le C\sum_{\ell=1}^dk_{h\ell}^{-2\alpha_\ell-1}\bbI\{k_{h\ell}<p_\ell/c_\ell\}+C\frac{V(\k_h)}{n},
        \end{equation}
        where $\c=(c_1,c_2,\dots,c_d)\transpose$ is defined in Assumption \ref{assume:cov-local} (ii). Moreover, 
        \begin{equation}
            \label{eq:minimax-Frobenius-upper-V}
                \inf_{\k_h}\sup_{\bfSigma\in\calV(\bfalpha,d,\epsilon,C)}p^{-1}\bbE\big\|\calL_h(\S_n;\k_h)-\bfSigma\big\|_F^2\le C\varepsilon_{n,\p}^\prime,
        \end{equation}
        where $\varepsilon_{n,\p}^\prime \asymp \underset{\k\in\calH_d(\p)}{\min}\big(\sum_{\ell=1}^dk_\ell^{-2\alpha_\ell-1}\bbI\{k_\ell<p_\ell\}+n^{-1}\prod_{\ell=1}^dk_\ell\big).$
\end{theorem}

Theorem \ref{thm:minimax-Frobenius-upper-V} demonstrates that the Frobenius risk of the localization estimator is controlled by a polynomially decayed term $\sum_{\ell=1}^dk_{h\ell}^{-2\alpha_\ell-1}\bbI\{k_{h\ell}<p_\ell/c_\ell\}$ and $V(\k_h)/n$. The former term on the right-hand side of \eqref{eq:minimax-Frobenius-upper-V-0} converges to $0$ if the scaling parameters $k_{h\ell}\to\infty$ as $n$ and $p_\ell\to\infty$, while the latter is guaranteed to vanish for $V(\k_h)=o(n)$. The upper bound $\varepsilon_{n,\p}^\prime$ in \eqref{eq:minimax-Frobenius-upper-V} is tighter than the upper bound term $\varepsilon_{n,\p}$ in \eqref{eq:minimax-spectral-upper-2} with $\tau(\k)$ given by \eqref{eq:multi-poly-cov}. The reason is that $\varepsilon_{n,\p}$ directly controls the bias of the localization estimator by $\tau^2(\k_h\circ\c)$, which can be improved as $\sum_{\ell=1}^dk_\ell^{-2\alpha_\ell-1}\bbI\{k_\ell<p_\ell/c_\ell\}$ for the Frobenius loss by averaging the bias of all the entries for the specific case $\bfSigma\in\calV(\bfalpha,d,\epsilon,C)$. Specifically, if $p_\ell$ is sufficient large such that $p_\ell>n^{(2\alpha_\ell+1)^{-1}\{1+\sum_{\ell=1}^d(2\alpha_\ell+1)^{-1}\}^{-1}}$ for $\ell=1,\dots,d$, then $\varepsilon_{n,\p}^\prime=n^{-\{1+\sum_{\ell=1}^d(2\alpha_\ell+1)^{-1}\}^{-1}}$, which is decided by the sample size $n$ and each $\alpha_\ell$. On the other hand, such an upper bound can be loose if the separable or multi-separable covariance structure is assumed, since a tighter bound may be attained under separability as given in \cite{Zhang2023} for $d=2$.

\begin{remark}
    Although the GC function does not satisfy Assumption \ref{assume:cov-local} (ii), for covariance matrices $\bfSigma\in\calV(\bfalpha,d,\epsilon,C)$, the localization estimator with the localization function being $\text{GC}(2\|\z\|)$ can attain the upper bound of the estimation error under the spectral norm \eqref{eq:minimax-spectral-upper-2} if $0<\alpha_\ell<2$ for $\ell=1,\dots,d$, and the upper bound of the estimation error under the Frobenius norm \eqref{eq:minimax-Frobenius-upper-V} if $0<\alpha_\ell<3/2$ for $\ell=1,\dots,d$. The key is to control the bias $\|\calL_{\text{GC}}(\S_n;\k_{\text{GC}})-\bfSigma\|\le C\sum_{\ell=1}^dk_{\text{GC},\ell}^{-\alpha_\ell}$ and $p^{-1}\|\calL_{\text{GC}}(\S_n;\k_{\text{GC}})-\bfSigma\|_F^2\le C\sum_{\ell=1}^dk_{\text{GC},\ell}^{-2\alpha_\ell-1}$. See the discussion in Section \ref{sec:discussion-GC} of the SM.
\end{remark}

We next study the minimax properties of the covariance estimation of tensor data within specific covariance classes. The following theorem demonstrates that the minimax upper bound under the spectral norm in \eqref{eq:minimax-spectral-upper-2} can not be further improved for the heterogeneous covariance decay function $\tau(k)=C\sum_{\ell=1}^d\big[k_\ell^{-\alpha_\ell}\bbI\{0< k_\ell<p_\ell\}+\bbI\{k_\ell=0\}\big]$ in \eqref{eq:multi-poly-cov}. 
\begin{theorem}
    \label{thm:minimax-spectral-lower}
    For Gaussian distributed $\X_1$, under Assumption \ref{assume:cov-decay}, for covariance decay function $\tau(\k)$ in \eqref{eq:multi-poly-cov}, if $\log p =o(n)$ and $p_\ell>n^{\alpha_\ell^{-1}(2+\sum_{\ell=1}^d\alpha_\ell^{-1})^{-1}}$ for $\ell=1,\dots,d$, the minimax risk of estimating the covariance matrix $\bfSigma\in\calU(d,\tau,\epsilon)$ satisfies
    \begin{equation}
        \label{eq:minimax-spectral-lower}
        \inf_{\hat\bfSigma}\sup_{\bfSigma\in\calU(d,\tau,\epsilon)}\bbE\big\|\hat\bfSigma-\bfSigma\big\|^2 \ge C\varepsilon_{n,\p}+C\frac{\log p}{n},
    \end{equation}
    where $\varepsilon_{n,\p}=\min_{\k\in\calH_d(\p)}\{\tau^2(\k)+V(\k)/n\}$. 
\end{theorem}
Then, according to Theorems \ref{thm:minimax-spectral-upper} and \ref{thm:minimax-spectral-lower}, it follows that
\begin{equation*}
    \inf_{\hat\bfSigma}\sup_{\bfSigma\in\calU(d,\tau,\epsilon)}\bbE\big\|\hat\bfSigma-\bfSigma\big\|^2 \asymp \varepsilon_{n,\p}+\frac{\log p}{n},
\end{equation*}
for the heterogeneous variance decay $\tau(\k)$ in \eqref{eq:multi-poly-cov}, $\log p =o(n)$ and $p_\ell>n^{\alpha_\ell^{-1}(2+\sum_{\ell=1}^d\alpha_\ell^{-1})^{-1}}$ for $\ell=1,\dots,d$. Hence, the localization estimator $\calL_h(\S_n;\k_h)$ in \eqref{eq:localization-estimator} is the minimax rate optimal under the spectral norm. 

As for the minimax optimal convergence rate under the Frobenius norm, we consider the following covariance class with homogeneous and polynomial decay
\begin{eqnarray*}
    \calW(\alpha,d,\epsilon,C)=\Big\{ \bfSigma=[\sigma_{ij}]_{p\times p}: &\text{(i)}& |\sigma_{ij}|\le C\|\bfdelta_{ij}\|^{-\alpha-d} \hbox{ for } \bfdelta_{ij}\neq\0_d, \\
   &\text{(ii)}& \add{0}\le \lambda_{\min}(\bfSigma)\le\lambda_{\max}(\bfSigma)\le \epsilon^{-1}\Big\}, \nn
\end{eqnarray*}
where $\alpha$ and $\epsilon$ are positive constants. One can see that $\calW(\alpha,d,\epsilon,C)\subset\calV(\bfalpha,d,\epsilon,C)$. The following theorem establishes the minimax lower bound of estimating $\bfSigma\in\calW(\alpha,d,\epsilon,C)$ under the Frobenius norm.
\begin{theorem}
    \label{thm:minimax-Frobenius-lower}
    For Gaussian distributed $\X_1$, under Assumption \ref{assume:cov-decay}, if $p_\ell>n^{1/(2\alpha+2d)}$ for all $\ell=1,\dots,d$, the minimax risk of estimating $\bfSigma\in\calW(\alpha,d,\epsilon,C)$ satisfies
    \begin{equation}
        \label{eq:minimax-Frobenius-lower-W}
        \inf_{\hat\bfSigma}\sup_{\bfSigma\in\calW(\alpha,d,\epsilon,C)}p^{-1}\bbE\big\|\hat\bfSigma-\bfSigma\big\|_F^2 \ge Cn^{-\frac{2\alpha+d}{2\alpha+2d}}.
    \end{equation}
\end{theorem}
Specifically, the localization estimator $\calL_h(\S_n;\k_h)$ in \eqref{eq:localization-estimator} with $h$ satisfying Assumption \ref{assume:cov-local} and $k_{h\ell}\asymp n^{1/(2\alpha+2d)}$ can attain the minimax rate of covergence specified in \eqref{eq:minimax-Frobenius-lower-W}. It then implies that the localization estimator $\calL_h(\S_n;\k_h)$ is the minimax rate optimal for estimating $\bfSigma\in\calW(\alpha,d,\epsilon,C)$ under the Frobenius norm. The detailed discussions for the minimax upper bounds and the general case when $\bfSigma\in\calV(\bfalpha,d,\epsilon,C)$ are provided in Section \ref{sec:class-W} of the SM.

\section{Simulation Study}
\label{sec:simulation}
This section reports results from simulation experiments designed to evaluate the performance of the proposed covariance localization estimator. To gain relative performance, the tapering estimator in \cite{Cai2010} for $d=1$ and the separably tapering estimator in \cite{Zhang2023} for $d=2$ were also considered.

The vectorized tensor data $\X_1,\X_2,\dots,\X_n$ were generated from Gaussian distributions and the $t_{10}$-distributions with zero mean in the numerical experiments. Three covariance structures were considered for the two distributions, which respectively had  
\begin{eqnarray*}
    (i) & & \sigma_{ij}=\sqrt{a_ia_j}\exp(-\|\s_i-\s_j\|^2/2);\\
    (ii) & & \sigma_{ij}=2\{(\lfloor (i-1)/p_1\rfloor+1)/p_2\}^{-|i-j|}\bbI\{\lfloor i-1/p_1\rfloor = \lfloor j-1/p_1\rfloor\}\hbox{ and }\\
    (iii) & & \sigma_{ij}=\sqrt{a_ia_j}\prod_{\ell=1}^3|s_{i\ell}-s_{j\ell}|^{-\alpha_\ell-1},
\end{eqnarray*}
as entries of the covariance matrix $\bfSigma=[\sigma_{ij}]_{p \times p}$, where $\{a_i\}_{i=1}^p$ were randomly drawn from a uniform distribution $\text{Unif}(0.5, 1.5)$ and were kept fixed once generated. The first structure was a homogeneous and $L_2$ distance-decayed covariance matrix and we considered $p=64,729$ and $4096$ grids in the $1$-, $2$- and $3$-order lattices, that is, the side lengths of the $1$-, $2$- and $3$-order lattices were set to be $p_1=64, 729, 4096$, $p_1=p_2=8, 81, 64$ and $p_1=p_2=p_3=4, 9, 16$, respectively. The second design was a block diagonal covariance matrix for $d=2$ that included a total number of $p_2$ block matrix of size $p_1$ with the entries of the $k$th block being $\sigma_{ij}=(k/p_2)^{-|i-j|}$, where $(p_1,p_2)$ was considered as $(10, 20)$, $(10, 50)$ and $(20, 50)$, respectively. The third one prescribed a heterogeneous setting for $d=3$ where we assigned $p_1=p_2=p_3=10$ and $(\alpha_1, \alpha_2,\alpha_3)=(0.4,0.6,0.8)$. The sample size ranged from $50$ to $2000$ and each experiment was replicated $500$ times. 

We employed the localization function $h(\z)=\prod_{\ell=1}^d\varphi(z_\ell;0.5,1)$, where $\varphi$ is the tapering function in \eqref{eq:linear-taper}. The scaling vector $\k_h$ for the proposed localization estimator was selected by the sample splitting scheme introduced in \cite{Bickel2008}. Specifically, a sample consisting of $n$ observations was randomly split into two subsamples of sizes $n_1=\lfloor n/3\rfloor$ and $n_2=n-n_1$, and the scaling vector for the localization estimator was chosen as
\begin{eqnarray}
    \label{eq:data-splitting}
    \hat\k_h=\arg\min_\k\sum_{b=1}^N\Big\|\calL_h\big(\S_{n,b}^{(1)};\k\big)-\S_{n,b}^{(2)}\Big\|_1,
\end{eqnarray}
where $\|\cdot\|_1$ denotes the matrix $L_1$ norm, $\S_{n,b}^{(u)}$ denotes the sample covariance matrix estimated by the $u$-th's split ($u=1$ or $2$) of the $b$-th simulated sample and $N$ was set to be $50$. The banding width for the tapering estimator in \cite{Cai2010} and the separably tapering estimator in \cite{Zhang2023} were selected similarly. 

\begin{table}[ht!]
    \centering
    \caption{Average empirical estimation errors and their standard deviations (in parentheses) of the proposed localization covariance estimator under the spectral and the Frobenius norms for Covariance Setting (i) for the Gaussian distributed  and the $t_{10}$-distributed data with respect to the dimension $p$, the sample size $n$ and the order of the lattice $d$.}
    \resizebox{\linewidth}{!}{ 
        \begin{tabular}{rrllllllll}
            \hline
            \multirow{3}{*}{$p$} & \multirow{3}{*}{$n$} &  & \multicolumn{7}{c}{Gaussian distribution} \\ 
            \cline{4-10}
             &  &  & \multicolumn{3}{c}{$\|\hat\bfSigma-\bfSigma\|$} & & \multicolumn{3}{c}{$\|\hat\bfSigma-\bfSigma\|_F$} \\ 
            \cline{4-6}\cline{8-10}
            & & & $d=1$ & $d=2$ & $d=3$ & & $d=1$ & $d=2$ & $d=3$ \\
            \cline{1-2}\cline{4-6}\cline{8-10}
            \multirow{6}{*}{64} & 50 &  & 1.07(0.24) & 1.88(0.3) & 2.95(0.52) &  & 2.92(0.26) & 4.4(0.37) & 5.83(0.57) \\ 
            & 100 &  & 0.77(0.13) & 1.48(0.22) & 2.37(0.4) &  & 2.03(0.22) & 3.45(0.25) & 4.56(0.36) \\ 
            & 250 &  & 0.5(0.09) & 0.97(0.17) & 1.7(0.33) &  & 1.33(0.12) & 2.16(0.23) & 3.25(0.3) \\ 
            & 500 &  & 0.37(0.06) & 0.73(0.13) & 1.2(0.26) &  & 1(0.08) & 1.6(0.13) & 2.24(0.26) \\ 
            & 1000 &  & 0.28(0.05) & 0.58(0.1) & 0.94(0.21) &  & 0.79(0.07) & 1.27(0.09) & 1.69(0.15) \\ 
            & 2000 &  & 0.19(0.04) & 0.48(0.09) & 0.78(0.16) &  & 0.48(0.04) & 1.05(0.09) & 1.39(0.11) \\
            \cline{1-2}\cline{4-6}\cline{8-10}
            \multirow{6}{*}{729} & 50 &  & 1.47(0.21) & 2.58(0.34) & 5.06(0.35) &  & 10.01(0.26) & 16.11(0.41) & 23.7(0.66) \\ 
            & 100 &  & 1.08(0.15) & 2(0.15) & 4.46(0.28) &  & 6.79(0.2) & 13.1(0.27) & 19.15(0.4) \\ 
            & 250 &  & 0.68(0.08) & 1.35(0.14) & 3.47(0.32) &  & 4.49(0.12) & 7.79(0.19) & 14.54(0.49) \\ 
            & 500 &  & 0.49(0.06) & 1.03(0.1) & 2.28(0.22) &  & 3.4(0.08) & 5.94(0.14) & 9.42(0.23) \\ 
            & 1000 &  & 0.36(0.04) & 0.84(0.07) & 1.96(0.16) &  & 2.71(0.06) & 4.77(0.11) & 7.41(0.17) \\ 
            & 2000 &  & 0.26(0.03) & 0.71(0.05) & 1.75(0.12) &  & 1.66(0.05) & 4.06(0.08) & 6.15(0.14) \\
            \cline{1-2}\cline{4-6}\cline{8-10}
            \multirow{6}{*}{4096} & 50 &  & 1.78(0.22) & 2.97(0.33) & 5.86(0.24) &  & 23.67(0.25) & 38.83(0.4) & 59.64(0.66) \\ 
            & 100 &  & 1.29(0.16) & 2.22(0.15) & 5.23(0.15) &  & 16.04(0.19) & 31.68(0.24) & 48.42(0.42) \\ 
            & 250 &  & 0.78(0.08) & 1.53(0.11) & 4.18(0.12) &  & 10.64(0.12) & 18.83(0.2) & 37.39(0.24) \\ 
            & 500 &  & 0.56(0.06) & 1.17(0.08) & 2.74(0.15) &  & 8.08(0.08) & 14.39(0.14) & 24.09(0.24) \\ 
            & 1000 &  & 0.41(0.03) & 0.94(0.06) & 2.34(0.1) &  & 6.44(0.06) & 11.57(0.1) & 18.96(0.18) \\ 
            & 2000 &  & 0.3(0.03) & 0.78(0.04) & 2.1(0.07) &  & 3.94(0.04) & 9.86(0.08) & 15.79(0.14) \\ 
            \hline
            \hline
            \multirow{3}{*}{$p$} & \multirow{3}{*}{$n$} &  & \multicolumn{7}{c}{$t$ distribution} \\ 
            \cline{4-10}
             &  &  & \multicolumn{3}{c}{$\|\hat\bfSigma-\bfSigma\|$} & & \multicolumn{3}{c}{$\|\hat\bfSigma-\bfSigma\|_F$} \\ 
            \cline{4-6}\cline{8-10}
            & & & $d=1$ & $d=2$ & $d=3$ & & $d=1$ & $d=2$ & $d=3$ \\
            \cline{1-2}\cline{4-6}\cline{8-10}
            \multirow{6}{*}{64} & 50 &  & 1.71(0.39) & 2.71(0.67) & 4.07(1.17) &  & 4.38(0.54) & 6.21(0.93) & 8.05(1.43) \\ 
            & 100 &  & 1.42(0.28) & 2.16(0.48) & 3.24(0.87) &  & 3.63(0.4) & 5.17(0.64) & 6.61(0.98) \\ 
            & 250 &  & 1.09(0.17) & 1.85(0.35) & 2.76(0.7) &  & 3.08(0.28) & 4.19(0.45) & 5.43(0.66) \\ 
            & 500 &  & 0.93(0.13) & 1.58(0.23) & 2.49(0.44) &  & 2.9(0.22) & 3.78(0.36) & 4.7(0.5) \\ 
            & 1000 &  & 0.85(0.11) & 1.4(0.18) & 2.19(0.35) &  & 2.82(0.17) & 3.58(0.26) & 4.32(0.39) \\ 
            & 2000 &  & 0.84(0.08) & 1.31(0.18) & 1.99(0.29) &  & 2.78(0.14) & 3.49(0.22) & 4.13(0.31) \\ 
            \cline{1-2}\cline{4-6}\cline{8-10}
            \multirow{6}{*}{729} & 50 &  & 2.36(0.34) & 3.88(0.55) & 6.34(0.86) &  & 14.93(0.59) & 22.21(0.98) & 31.56(1.67) \\ 
            & 100 &  & 1.91(0.23) & 2.87(0.33) & 4.74(0.56) &  & 12.3(0.45) & 18.93(0.7) & 26.29(1.19) \\ 
            & 250 &  & 1.4(0.13) & 2.52(0.26) & 3.91(0.47) &  & 10.52(0.29) & 14.79(0.51) & 21.71(0.75) \\ 
            & 500 &  & 1.16(0.1) & 2.05(0.17) & 3.73(0.32) &  & 9.85(0.22) & 13.39(0.37) & 17.99(0.59) \\ 
            & 1000 &  & 0.98(0.07) & 1.73(0.12) & 3.11(0.23) &  & 9.51(0.18) & 12.64(0.29) & 16.43(0.46) \\ 
            & 2000 &  & 0.98(0.06) & 1.52(0.09) & 2.71(0.17) &  & 9.43(0.14) & 12.26(0.22) & 15.63(0.33) \\ 
            \cline{1-2}\cline{4-6}\cline{8-10}
            \multirow{6}{*}{4096} & 50 &  & 2.76(0.31) & 4.56(0.46) & 7.65(0.78) &  & 35.45(0.58) & 53.46(0.94) & 78.59(1.54) \\ 
            & 100 &  & 2.23(0.22) & 3.33(0.31) & 5.54(0.52) &  & 29.11(0.43) & 45.48(0.69) & 65.41(1.17) \\ 
            & 250 &  & 1.58(0.12) & 2.86(0.21) & 4.35(0.29) &  & 24.93(0.29) & 35.5(0.5) & 54.33(0.8) \\ 
            & 500 &  & 1.28(0.09) & 2.27(0.15) & 4.28(0.27) &  & 23.37(0.21) & 32.12(0.36) & 44.65(0.63) \\ 
            & 1000 &  & 1.07(0.06) & 1.9(0.11) & 3.51(0.18) &  & 22.55(0.17) & 30.25(0.29) & 40.56(0.49) \\ 
            & 2000 &  & 1.05(0.05) & 1.65(0.08) & 2.97(0.13) &  & 22.33(0.14) & 29.28(0.22) & 38.36(0.37) \\ 
            \hline
        \end{tabular}
        \label{tab:case1}
    }
\end{table}

Table \ref{tab:case1} summarizes the empirical estimation errors of the proposed localization estimator for Covariance Setting (i) under both the spectral and Frobenius norms with respect to the dimension $p$, the sample size $n$ and the order of the lattice $d$. It shows that for each combination of dimension $p$ and order of lattice $d$, the estimation errors under both the spectral and Frobenius norms decreased as the sample size $n$ increased. In the meanwhile, a similar trend happened for the standard deviations of the average empirical errors, indicating the variation of the estimation errors was reduced along with the increase of the sample size. When the dimension $p$ and sample size $n$ were fixed, both the spectral norm and the Frobenius norm of the estimation errors increased as the order of the lattice $d$ got larger, which was consistent with the discussion in Section \ref{sec:main-result} that larger $d$ leads to a slower rate of convergence. Besides, the estimation errors for the samples generated from the $t$-distributions were larger than those for the Gaussian distribution under either the spectral norm or the Frobenius norm and for each dimension $p$, samples size $n$ and order of the lattice $d$, which reflected the fact that the $t$-distribution has heavier tail than the normal distribution. In the meanwhile, simulation results of the proposed localization covariance estimator using the multiplicative banding function $h(\z)=\prod_{\ell=1}^d\bbI(z_\ell<1)$ and the GC function $h(\z)=\text{GC}(2\|\z\|)$ are reported in Tables \ref{tab:case1-banding} and \ref{tab:case1-GC} in Section  \ref{sec:add-experiment} of the SM, which show similar patterns of results as those in Table \ref{tab:case1}. 

\begin{table}[ht!]
    \centering
    \caption{Average empirical estimation errors and their standard deviations (in parentheses) ofthe localization estimator (proposed), the sample covariance  (sample) and the tapering estimator (CZZ), the separably tapering estimator (ZSK) under the spectral and the Frobenius norms for Covariance Setting (ii) for $d=2$ for the Gaussian distributed data with respect to the dimension of the data $p$ and the sample size $n$. (Zero standard deviations indicate values below $0.005$)}
    \resizebox{\linewidth}{!}{
        \begin{tabular}{lllllllllll}
            \hline
            \multirow{2}{*}{n} & & \multicolumn{4}{c}{$\|\hat\bfSigma-\bfSigma\|$} & & \multicolumn{4}{c}{$\|\hat\bfSigma-\bfSigma\|_F$}\\
            \cline{3-11}
            ~ & & sample & CZZ & ZSK & proposed & & sample & CZZ & ZSK & proposed\\ 
            \hline
             & & \multicolumn{9}{c}{$(p_1,p_2)=(10,20)$} \\
            \cline{3-11}
            50 &  & 21.27(0.11) & 8.39(0.1) & 7.71(0.1) & 6.56(0.1) &  & 57.52(0.07) & 20.05(0.12) & 20.82(0.08) & 16.11(0.13) \\ 
            100 &  & 13.65(0.07) & 5.8(0.08) & 6.36(0.06) & 4.37(0.08) &  & 40.49(0.04) & 14.27(0.09) & 19.24(0.05) & 11(0.1) \\ 
            250 &  & 7.89(0.04) & 3.26(0.04) & 5.51(0.02) & 2.39(0.03) &  & 25.49(0.02) & 8.71(0.04) & 18.31(0.01) & 6.4(0.03) \\ 
            500 &  & 5.42(0.02) & 2.33(0.03) & 5.38(0.01) & 1.71(0.03) &  & 18.04(0.01) & 6.26(0.02) & 18.14(0) & 4.53(0.02) \\ 
            1000 &  & 3.7(0.01) & 1.56(0.02) & 5.32(0.01) & 1.18(0.02) &  & 12.74(0.01) & 4.45(0.02) & 18.04(0) & 3.18(0.01) \\ 
            2000 &  & 2.58(0.01) & 1.06(0.01) & 5.33(0.01) & 0.81(0.01) &  & 8.99(0.01) & 3.18(0.01) & 17.99(0) & 2.23(0.01) \\  
            \hline
            & & \multicolumn{9}{c}{$(p_1,p_2)=(10,50)$} \\
            \cline{3-11}
            50 &  & 41.1(0.12) & 9.28(0.1) & 8.46(0.1) & 7.76(0.11) &  & 143.24(0.1) & 31.41(0.21) & 32.11(0.15) & 25.76(0.22) \\ 
            100 &  & 25.5(0.07) & 6.67(0.08) & 7.18(0.08) & 5.47(0.09) &  & 100.79(0.05) & 22.8(0.17) & 29.97(0.1) & 18.34(0.21) \\ 
            250 &  & 14.15(0.04) & 3.96(0.05) & 6.03(0.03) & 3.08(0.05) &  & 63.53(0.02) & 14.1(0.07) & 28.26(0.03) & 10.63(0.08) \\ 
            500 &  & 9.39(0.03) & 2.61(0.03) & 5.69(0.01) & 1.96(0.02) &  & 44.87(0.01) & 9.86(0.04) & 27.85(0) & 7.11(0.02) \\ 
            1000 &  & 6.32(0.02) & 1.89(0.02) & 5.65(0.01) & 1.38(0.02) &  & 31.71(0.01) & 7.14(0.02) & 27.72(0) & 4.99(0.01) \\ 
            2000 &  & 4.33(0.01) & 1.28(0.01) & 5.63(0.01) & 1(0.01) &  & 22.42(0.01) & 5.06(0.02) & 27.67(0) & 3.55(0.01) \\
            \hline
            & & \multicolumn{9}{c}{$(p_1,p_2)=(20,50)$} \\
            \cline{3-11}
            50 &  & 76.67(0.23) & 21.74(0.23) & 19.13(0.26) & 16.61(0.27) &  & 286.29(0.16) & 58.18(0.32) & 57.99(0.27) & 49.39(0.37) \\ 
            100 &  & 46.91(0.14) & 16.52(0.2) & 16.35(0.2) & 11.86(0.24) &  & 201.34(0.08) & 44.47(0.28) & 54.21(0.17) & 35.7(0.34) \\ 
            250 &  & 26.07(0.07) & 9.1(0.14) & 12.78(0.08) & 5.63(0.11) &  & 126.95(0.03) & 27.34(0.16) & 50.99(0.05) & 20.25(0.14) \\ 
            500 &  & 17.3(0.05) & 5.64(0.08) & 11.99(0.03) & 3.51(0.05) &  & 89.66(0.02) & 18.85(0.06) & 50.4(0.01) & 13.64(0.03) \\ 
            1000 &  & 11.63(0.03) & 3.69(0.04) & 11.71(0.02) & 2.5(0.03) &  & 63.39(0.01) & 13.38(0.03) & 50.25(0) & 9.66(0.02) \\ 
            2000 &  & 7.93(0.02) & 2.48(0.03) & 11.6(0.01) & 1.75(0.02) &  & 44.81(0.01) & 9.63(0.02) & 50.16(0) & 6.82(0.01) \\ 
            \hline
        \end{tabular}
    }
    \label{tab:case2}
\end{table}

\begin{figure}[ht!]
    \centering
    \includegraphics[width=\linewidth]{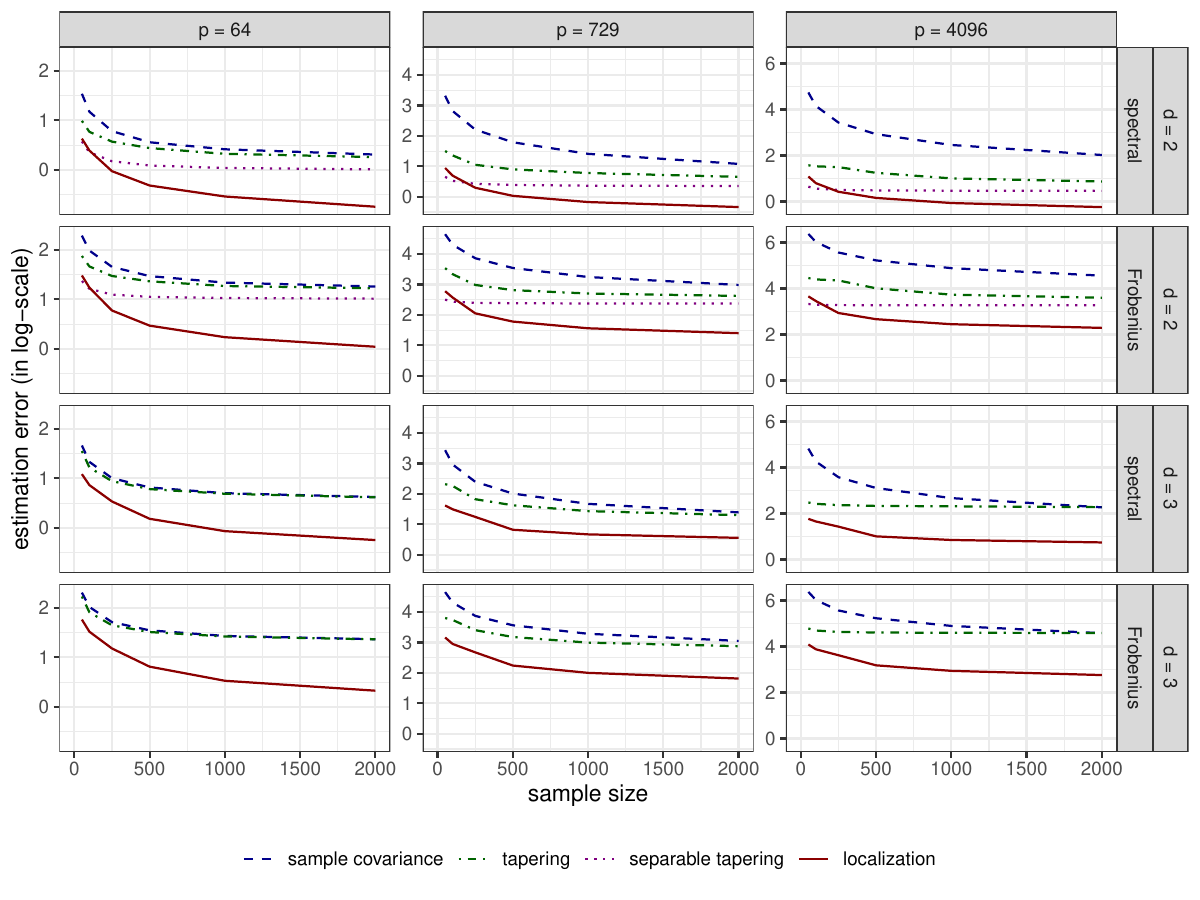}
    \caption{Average empirical estimation errors (in log-scale) of the proposed localization estimator (red solid lines), the sample covariance (blue dashed lines), the tapering estimator (green dashed-dotted lines) for $d=2$ and $3$, and the separably tapering estimator (purple dotted lines) (for $d=2$ only)  under the spectral and Frobenius norms with respect to the dimension $p$ and the sample size $n$ for Covariance Setting (i) of the Gaussian-distributed data.}
    \label{fig:higherD}
\end{figure}

Figure \ref{fig:higherD} displayed the average empirical estimation errors under the spectral and Frobenius norms of the proposed localization estimator with the sample covariance for $d=2$ and $3$, the tapering estimator for $d=2,3$ and the separably tapering estimator for $d=2$ for the Gaussian-distributed data. The figure shows that the proposed localization estimator had decreasing estimation errors as the sample size $n$ increased and achieved the smallest estimation errors in most settings for $d=2$ and all the settings for $d=3$. In contrast, the sample covariance generally obtained the largest estimation errors, and the estimation errors of the tapering estimator were the secondly largest among the four covariance matrix estimators. The tapering estimator designed for $d=1$ had very subdued decreases in the estimation error as the sample size increased. The latter might be due to the tapering estimator designed for recovering univariate bandable covariance structure was too general to capture the more detailed structure of the covariance matrix for multi-order lattices and could not capture the multi-bandable structure as displayed in Figure \ref{fig:hyper-rectangle}, which resulted in excessive bias. The separably tapering estimator \eqref{eq:ZSK-approximation} obtained relatively smaller estimation errors for $d=2$ when the sample size was quite small. The reasons can be attributed to the fact that the non-separability of Covariance Setting (i) is from the randomly drawn $\{a_i\}$ only, hence the variance reduction from the approximation \eqref{eq:ZSK-approximation} could offset the additional introduced bias. However, less improvement could be obtained for the estimation errors in Covariance Setting (i) as the sample size increased, which yields that solving \eqref{eq:ZSK-approximation} to approximate the separably tapering estimator can sometimes be harmful to the non-separable scenarios.

We further compared the localization estimator with the tapering or the separably tapering estimators in a more heterogeneous case of Covariance Setting (ii), where the decay pattern of the covariance in each block is completely different. Table \ref{tab:case2} reports the average empirical estimation errors of the proposed localization estimator for Gaussian distributed data under both the spectral and the Frobenius norms, as well as the results of the sample covariance matrix, the tapering and the separably tapering estimators. Although the estimation errors under both the spectral and the Frobenius norms of all the estimators saw decreases as either the dimensions $p_1$ and $p_2$ decreased or the sample size $n$ increased, the proposed localization estimator outperformed in all the settings with respect to difference dimensions $p$ and sample sizes $n$ and obtained the smallest estimation errors compared with the sample covariance matrix, the tapering estimator and the separably tapering estimator. In particular, the separably tapering estimator obtained the largest estimation errors under the spectral and Frobenius norms for $n=2000$ as it tended to construct the same covariance pattern among all the blocks in Covariance Setting (ii), which introduced non-negligible bias. It also demonstrated that the localization estimator may be a sound choice if the separability is not guaranteed. 

\begin{table}[ht!]
    \centering
    \caption{Average empirical estimation errors of the proposed localization estimator under the spectral and the Frobenius norms and selected scaling parameters with their standard deviations in the parentheses for Covariance Setting (iii) for $d=3$ for the Gaussian distributed and the $t_{10}$-distributed data with respect to the dimension of the data $p_1=p_2=p_3=10$ and the sample size $n$ from $50$ to $2000$. (Zero standard deviations indicate values below $0.005$)}
    \begin{tabular}{llllllll}
        \hline
        n & & $\|\hat\bfSigma-\bfSigma\|$ & $\|\hat\bfSigma-\bfSigma\|_F$ & &  $k_1$ & $k_2$ & $k_3$ \\ 
        \hline
        & & \multicolumn{6}{c}{Gaussian distribution} \\ 
        \cline{3-8}
        50 & & 27.77(1.98) & 59.84(1.82) & & 1.05(0.22) & 3(0) & 6.31(0.72) \\ 
        100 & & 20.11(1.66) & 45.14(1.67) & & 2(0) & 3(0) & 5.83(0.77) \\ 
        250 & & 15.43(1.5) & 34.44(1.19) & & 2.37(0.48) & 3.57(0.5) & 7.43(0.62) \\ 
        500 & & 10.8(1.2) & 25.1(1.04) & & 2.98(0.13) & 4.37(0.48) & 8.88(0.32) \\ 
        1000 & & 8.93(0.73) & 19.56(0.51) & & 3(0) & 5(0) & 9.81(0.39) \\ 
        2000 & & 7.09(0.66) & 15.51(0.48) & & 3(0) & 5.92(0.27) & 11(0) \\ 
        \cline{3-8}
        & & \multicolumn{6}{c}{$t$ distribution} \\ 
        \cline{3-8}
        50 & & 25.05(2.21) & 64.92(1.53) & & 1.05(0.22) & 3(0) & 6.22(0.76) \\ 
        100 & & 16.09(1.9) & 52.29(1.52) & & 2(0) & 3(0) & 5.75(0.77) \\ 
        250 & & 11.59(1.28) & 42.67(1.36) & & 2.32(0.47) & 3.53(0.51) & 7.52(0.68) \\ 
        500 & & 10.03(0.94) & 35.37(1.24) & & 2.95(0.22) & 4.4(0.49) & 8.83(0.37) \\ 
        1000 & & 9.1(0.72) & 30.75(1.03) & & 3(0) & 5(0) & 9.74(0.44) \\ 
        2000 & & 8.9(0.58) & 28.32(0.85) & & 3(0) & 5.83(0.37) & 11(0.06) \\ 
        \hline
    \end{tabular}
    \label{tab:case3}
\end{table}

Table \ref{tab:case3} reports the average estimation errors of the proposed localization estimator and the selected localization scaling parameters for Covariance Setting (iii). In general, the estimation errors under both the spectral and the Frobenius norms saw significant declines as the sample size $n$ increased for both the Gaussian-distributed data and the t-distributed data. For a sample size $n$, the $t$-distributed data had larger estimation errors and standard deviations than the Gaussian-distributed data under both the spectral and the Frobenius norms. The chosen scaling parameters $k_1,k_2,k_3$ were close in average between the Gaussian-distributed and the t-distributed data, while the corresponding standard deviations were in general larger for the t-distributed data except for $k_1$ for $n=250$. As the sample size $n$ increased, the selected scaling parameters $k_1$, $k_2$ and $k_3$ saw increasing as pointed out in the discussion of Section \ref{sec:main-result}. Among all the scaling parameters, $k_3$ was the largest for each sample size $n$ for both the Gaussian-distributed and the t-distributed data, which corresponds to $\alpha_3=0.8$, the direction with the slowest decay rate. The above results demonstrate that the proposed localization estimator is suitable for the heterogeneous case when the decay patterns in each direction of the lattice are different by properly choosing the localization scaling parameters.

\section{Case Study}
\label{sec:case}
Oceanic eddies play a critical role in modulating ocean heat exchange, nutrient distribution and climate variability \citep{Xu2016, Beech2022, He2024, Receveur2024}. Reconstruction of the salinity fields of an ocean eddy helps to provide high-resolution insights into ocean dynamics and enhance oceanographic studies. We analyzed in this section the covariance matrix of the daily salinity changes of the eddy data introduced in Section \ref{sec:setting}, which was the reanalysis data from GLORYS from July 20th to September 12th (54 days), 2024. We re-centered the eddy each day and calculated the salinity changes between two consecutive days so that each daily salinity change was treated as a replication. The state variable consisted of daily salinity changes over $p=47915$ grids, which was a result of $37$ longitude and latitude grid partitions and $35$ vertical divisions. A sample of $n=54$ observations on the daily changes was then obtained. Despite the eddy's salinity field was likely dependent over time. The daily changes should be much less dependent. We treat them as independent in our study. 

We considered the localization estimator $\calL_h(\S_n;\k_h)$ in \eqref{eq:localization-estimator} with the localization function being the multiplicative banding function $h(\z)=\prod_{\ell=1}^3\bbI\{z_\ell<1\}$ and the multiplicative tapering function $h(\z)=\prod_{\ell=1}^3\varphi(z_\ell;0.5,1)$. The scaling vector $\k_h=(k_{h1},k_{h2},k_{h3})\transpose$ of the localization estimator were selected via the data splitting procedure in \eqref{eq:data-splitting} with the number of replications being $N=50$ and were chosen from the set $\{0.1,0.2,\dots,1\}$ degree for the longitude and the latitude, and $\{10, 20,\dots, 200\}$ meter for the depth. Figures \ref{fig:scale-localization-banding} and \ref{fig:scale-localization-tapering} in the SM display the objective function in \eqref{eq:data-splitting} using the two localization functions for each combination of the scaling parameters. The optimal scaling parameters were chosen as $0.2^\circ$ in longitude, $0.3^\circ$ in latitude and $80\mathrm{m}$ in depth for the multiplicative banding function, and $0.3^\circ$ in longitude, $0.4^\circ$ in latitude and $200\mathrm{m}$ in depth for the multiplicative tapering function. We would like to put the chosen parameter in the perspective of oceanography. The chosen scaling parameters in the longitude and the latitude were around $30\mathrm{km}$, which contained the first baroclinic Rossby radius of deformation, an important quantity in determining horizontal scales that characterize eddy sizes \citep{Chelton1998}. 

\begin{figure}[ht!]
    \centering
    \includegraphics[width=\linewidth]{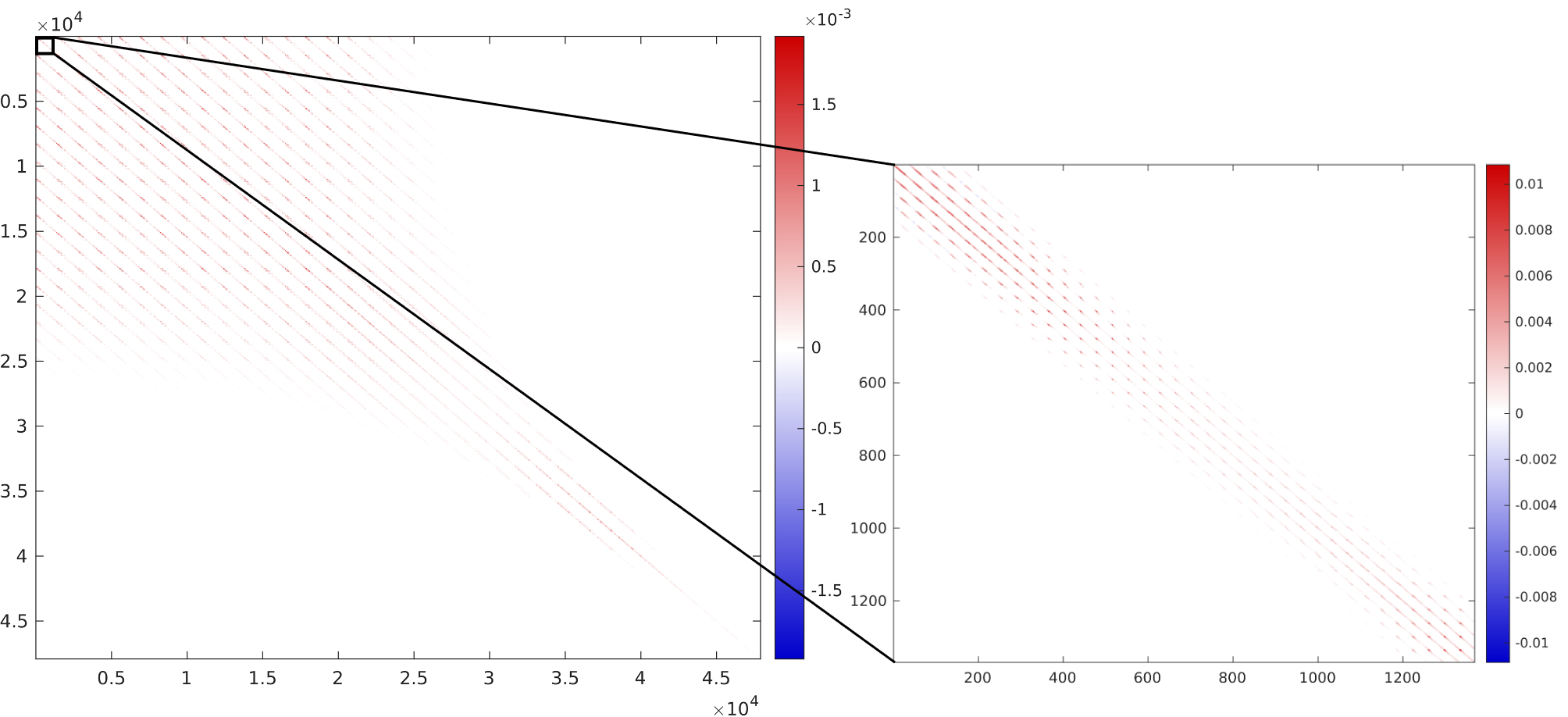}
    \caption{Localization estimator with the localization function being the multiplicative banding function and insert of the covariance matrix of daily salinity changes in the ocean eddy region.}
    \label{fig:eddy-local}
\end{figure}

Figure \ref{fig:eddy-local} illustrates the proposed covariance localization estimator with the localization function being the multiplicative banding function and an insert corresponding to the localization estimator of the sea surface of the daily salinity changes. Compared to the sample correlation matrix in Figure \ref{fig:Eddy-2}, a finer multi-bandable covariance structure for the trivariate ocean tensor data was successfully captured. The precision of the covariance matrix estimation was verified via a three-dimensional variational assimilation (3DVar) framework, which critically depended on the quality of estimation on a key covariance matrix. To be specific, we estimated the covariance matrix $\bfSigma$ of the daily salinity changes using the daily changes in salinity over the first $53$ days, while the data on September 12th, 2024, the last date of the study period, was used as the testing set. We randomly selected the salinity data on $5\%$ of the grids while adding a $\calN(0,0.01\I)$ distributed noise to the observation on each observed grid. The aim was to reconstruct the salinity field on the rest $95\%$ grids. Denote by $\Y$ the observations on the observed grids and let $\H$ be a matrix that maps the state variables to the observations. The 3DVar assimilated the salinity field on the last day as 
\begin{equation}
    \label{eq:KF-update}
    \hat\X=\X_0+\hat\bfSigma\H\transpose\big(\H\hat\bfSigma\H\transpose+\R\big)^{-1}(\Y-\H\X),
\end{equation}
where $\hat{\X}$ was the estimated salinity changes in the last day, $\X_0$ was a first guess on the increment which was set to be $0$ psu as it reflected the salinity anomaly, $\hat\bfSigma$ was the covariance estimation of the daily salinity changes using the first $53$ days' data and $\R=0.01\I$ represented the observational error covariance matrix. 

The estimate $\hat\bfSigma$ was obtained based on the localization estimator $\calL_h(\S_n;\k_h)$ employing the two localization functions $\prod_{\ell=1}^3\bbI\{z_\ell<1\}$ and $\prod_{\ell=1}^3\varphi(z_\ell;0.5,1)$, the sample covariance, the banding estimator $\calB_k(\S_n)$ in \eqref{eq:banding-estimator}, and the tapering estimator $\calT_{k}(\S_n)$ in \eqref{eq:linear-taper-estimator}. The scaling parameters of the localization estimator for the two functions were set to be the same as the aforementioned optimal scaling parameters selected via data splitting procedure in \eqref{eq:data-splitting}. The banding width parameters for the banding and tapering estimators were chosen as $3$ and $4$, respectively, via the data splitting procedure, whose objective function values can be seen in Figure \ref{fig:bandwidth-banding-tapering} of the SM. For each covariance matrix estimator, the 3DVar assimilation was replicated for $500$ times, in which the $5\%$ observed grids and the noises added to the observations collected in $\Y$ were randomly generated in each repetition. We treated the last day as the test data to verify the accuracy of the reconstruction on the $95\%$ missed salinity values on the last day.

\begin{table}[ht!]
    \centering
    \caption{Average reconstruction errors and their $5\%$ and $95\%$ quantiles in parentheses of the reconstructed state variable $\hat{\X}$ in the salinity field on the last day by using the sample covariance, the banding and tapering estimators and the localization estimators with multiplicative banding and tapering functions, including the $L_2$ distance, the $L_1$ distance and the Hamming distance divided by $p$.}
    \begin{tabular}{llll}
        \hline
        Estimation & $\|\X-\hat{\X}\|$ & $\|\X-\hat{\X}\|_1$ & $p^{-1}H\{\text{sgn}(\X),\text{sgn}(\hat{\X})\}$ \\ 
        \hline
        sample covariance & 6.14(6.11,6.19) & 918.5(905.97,932.76) & 0.35(0.34,0.36) \\ 
        banding & 7.5(7.47,7.52) & 1065.4(1063.1,1067.5) & 0.48(0.48,0.49) \\
        tapering & 7.48(7.46,7.51) & 1063.3(1061.1,1065.7) & 0.48(0.48,0.49) \\
        localization(banding) & 4.59(4.46,4.74) & 685.5(673.3,698.7) & 0.23(0.22,0.24) \\
        localization(tapering) & 4.46(4.33,4.58) & 664.5(652.4,676.5) & 0.21(0.21,0.22) \\
        \hline
    \end{tabular}
    \label{tab:eval}
\end{table}

Table \ref{tab:eval} summarizes the reconstruction errors on the $95\%$ ``missing" grids on the last day with different covariance estimators for $\hat\bfSigma$ being used in (\ref{eq:KF-update}). Three metrics were used in presenting the reconstruction errors, namely the $L_2$ distance $\|\X-\hat{\X}\|$, the $L_1$ distance $\|\X-\hat{\X}\|_1$ and the Hamming distance between $\text{sgn}(\X)$ and $\text{sgn}(\hat{\X})$, where $\text{sgn}(\X)$ is the sign indicator function. The results demonstrated that the reconstruction with the two localization covariance estimators $\calL_h(\S_n;\k_h)$ recovered the underlying salinity field more accurately with much smaller reconstruction errors in all three metrics than those using the sample covariance. In contrast, the banding and tapering estimators encountered larger reconstruction errors even than those using the sample covariance and this might be due to the two covariance estimators were designed for one-order lattice data, and were not suited for the three-order lattice data that we were dealing with here.

\section{Conclusion}
\label{sec:conclusion}
The central idea of this study is to explore the \hd\ covariance estimation of tensor data. For this purpose, the multi-bandable covariance class and the corresponding localization estimator are proposed to capture the refined covariance structure by regularizing the covariance estimations of two far-away grids with the localization functions. Theoretical analysis demonstrates a strong performance of the proposed approach with established minimax optimal rates of convergence under both spectral and Frobenius norms, which advances covariance estimation of tensor data and offers a scalable and adaptable framework for applications across scientific disciplines.

%%%%%%%%%%%%%%%%%%%%%%%%%%%%%%%%%%%%%%%%%%%%%%
%% Support information, if any,             %%
%% should be provided in the                %%
%% Acknowledgements section.                %%
%%%%%%%%%%%%%%%%%%%%%%%%%%%%%%%%%%%%%%%%%%%%%%
% \begin{acks}[Acknowledgments]
%     The authors would like to thank the anonymous referees, an Associate Editor and the Editor for their constructive comments that improved the quality of this paper.
% \end{acks}

%%%%%%%%%%%%%%%%%%%%%%%%%%%%%%%%%%%%%%%%%%%%%%
%% Funding information, if any,             %%
%% should be provided in the                %%
%% funding section.                         %%
%%%%%%%%%%%%%%%%%%%%%%%%%%%%%%%%%%%%%%%%%%%%%% 
% \begin{funding}
% This research was partially supported by the National Natural Science Foundation of China grants 12292980, 12292983 and 92358303.
% \end{funding}

%%%%%%%%%%%%%%%%%%%%%%%%%%%%%%%%%%%%%%%%%%%%%%
%% Supplementary Material, including data   %%
%% sets and code, should be provided in     %%
%% {supplement} environment with title      %%
%% and short description. It cannot be      %%
%% available exclusively as external link.  %%
%% All Supplementary Material must be       %%
%% available to the reader on Project       %%
%% Euclid with the published article.       %%
%%%%%%%%%%%%%%%%%%%%%%%%%%%%%%%%%%%%%%%%%%%%%%
\begin{supplement}
\stitle{Supplementary Material to ``Localization Estimator for High Dimensional Tensor Covariance Matrices"} 
\sdescription{In the supplementary material, we present technical details, proofs and additional results of the simulations and the case study.}
\end{supplement}

%%  MR numbers will be added by VTeX.                      %%
%%                                                         %%
%%  Use \cite{...} to cite references in text.             %%
%%                                                         %%
%%%%%%%%%%%%%%%%%%%%%%%%%%%%%%%%%%%%%%%%%%%%%%%%%%%%%%%%%%%%%

%% if your bibliography is in bibtex format, uncomment commands:
\bibliographystyle{imsart-nameyear} % Style BST file (imsart-number.bst or imsart-nameyear.bst)
\bibliography{reference}       % Bibliography file (usually '*.bib')

\end{document}